\documentstyle[12pt]{article}

\advance\voffset by -2.0cm
\advance\hoffset by -1.5cm
\input amssym.def
\input amssym.tex
\def\ben{\begin{equation}}
\def\een{\end{equation}}
\def\bea{\begin{eqnarray}}
\def\eea{\end{eqnarray}}
\def\nn{\nonumber \\ }
\def\={\stackrel{\rm def}{=}}

\def\Nop{{\Bbb N}}
\def\Zop{{\Bbb Z}}
\def\Qop{{\Bbb Q}}
\def\Rop{{\Bbb R}}
\def\Cop{{\Bbb C}}

\def\mod{\ {\rm mod}\ } 
\def\ker{\mathop{\rm ker}}
\def\hcf{{\rm hcf}}
\def\Dt{{\widetilde D}}
 \def\Ddt{{\widetilde {D}'}}
 \def\Dddt{{\widetilde {D}''}}

 \def\chir{{\raise 1.5pt\hbox{$\chi$}}}
 \def\con#1{F{#1}} 
\def\ah{{a}}
\def\uh{{\hat u}}
\def\LHS{{\rm LHS}}
\def\a{\alpha}
\def\b{\beta}
\def\g{\gamma}
\def\th{\theta}
\def\e{\epsilon}
\def\gal{{\rm Gal}}
\def\pmb#1{\setbox0=\hbox{#1}%
 \kern-.025em\copy0\kern-\wd0
 \kern.05em\copy0\kern-\wd0
 \kern-.025em\raise.0433em\box0 }

\def\3{\ss}
\def\sq{\hbox{\rlap{$\sqcap$}$\sqcup$}}
\def\qed{\ifmmode\sq\else{\unskip\nobreak\hfil
\penalty50\hskip1em\null\nobreak\hfil\sq
\parfillskip=0pt\finalhyphendemerits=0\endgraf}\fi}

\def\Ok{{{\cal O}_{{\bf{K}}}}}
\def\Okp{{{\cal O}_{{\bf{K}'}}}}
\def\f{{\bf K}}
\def\fp{{{\bf K}'}}
\def\fk0{{\bf K}_0}
\def\I{{\cal{I}}({\bf K})}
\def\P{{\cal{P}}({\bf K})}
\def\IF{{\cal{I}}_F({\bf K})}
\def\CF{{\cal{C}}_F({\bf K})}
\def\CI{{\cal{C}}_I({\bf K})}

\def\PF{{\cal{P}}_F({\bf K})}
\def\CPF{{\cal{CP}}_F({\bf K})}

\def\a{\alpha}
\def\r{\rho}
\def\l{\ell}

\def\De{\Delta}
\def\D0{\Delta_0}
\def\Dp{\Delta '}
\def\sqD{\sqrt{D}}
\def\sqDp{\sqrt{D'}}
\def\sqk{\sqrt{k}}
\def\sql{\sqrt{\ell }}
\def\eqalign#1{\vcenter{\halign{$\hfil##\strut$&$\strut##\hfil$\cr#1\crcr}}}
\def\rp{{\sqrt{-30}}}
\def\rpd{{\sqrt{-10}}}
\def\by{\times}
\def\Okpp{{{\cal O}_{{\bf{K}''}}}}
\def\Okkp{{{\cal O}_{{\bf KK'}}}}

\def\bbbc{{\mathchoice {\setbox0=\hbox{$\displaystyle\rm C$}\hbox{\hbox
to0pt{\kern0.4\wd0\vrule height0.9\ht0\hss}\box0}}
{\setbox0=\hbox{$\textstyle\rm C$}\hbox{\hbox
to0pt{\kern0.4\wd0\vrule height0.9\ht0\hss}\box0}}
{\setbox0=\hbox{$\scriptstyle\rm C$}\hbox{\hbox
to0pt{\kern0.4\wd0\vrule height0.9\ht0\hss}\box0}}
{\setbox0=\hbox{$\scriptscriptstyle\rm C$}\hbox{\hbox
to0pt{\kern0.4\wd0\vrule height0.9\ht0\hss}\box0}}}}

\newtheorem{Theorem}{Theorem}[section]
\newtheorem{Lemma}[Theorem]{Lemma}
\newtheorem{Definition}[Theorem]{Definition}
\newtheorem{Proposition}[Theorem]{Proposition}
\newtheorem{Corollary}[Theorem]{Corollary}
\begin{document}
\def\thelabel{}%
\let\oldlabel=\label
\def\label#1{\def\thelabel{#1}\oldlabel{#1}}
\catcode`@=11
\def\@eqnnum{{\rm(\theequation)}\def\thelabel{}}
\catcode`@=12
%
\def\theequation{\thesection.\arabic{equation}}
\begin{titlepage}
\begin{flushright}
DTP/97/23\\
May 1997\\
\end{flushright}
\vskip 2.cm
\begin{center}                                                                  
{\Large\bf Virasoro character identities and Artin L-functions.}
\vskip 2cm
{\large A. Taormina\footnote{email address Anne.Taormina@durham.ac.uk}
and \large S.M.J. Wilson
\footnote{email address S.M.J.Wilson@durham.ac.uk}}
\vskip 0.5cm
{\it Department of Mathematical Sciences, University of Durham, \\
Durham DH1 3LE, England }\\
\vskip 3cm                                                                    
\end{center}                                                                \begin{center}
{\bf ABSTRACT}
\end{center}
\begin{quote}
Some identities between unitary minimal Virasoro characters at 
levels $m=3,4,5$
are shown to arise as a consequence of relations between Artin 
L-functions of different quadratic fields. The definitions and concepts
of number theory necessary to present the theta function identities 
which can be derived from these relations are introduced.  
A new infinite family of identities between Virasoro characters 
at level 3
and level $m=4a^2$, for $a$ odd and $1+4a^2={a'}^2 p$ where $p$ is 
prime is obtained as a by-product.
\end{quote}
\vfill
\end{titlepage}

\section{Introduction}
\renewcommand{\theequation}{1.\arabic{equation}}
\setcounter{equation}{0}

Two sets of intriguing identities between unitary minimal Virasoro 
characters were presented in \cite{anne}. 
They are remarkable in various respects. First
they provide a rewriting of the Virasoro characters at level $m=3$ (resp.
$m=4$) in terms of $\em{differences}$ of quadratic expressions in Virasoro
characters at level $m=4$ (resp. $m=3,5$). Second, they are stronger 
identities
than the ones obtained by repeated use of the Goddard-Kent-Olive
sumrules \cite{GKO} for the cosets
$[SU(2)_1 \times SU(2)_1 \times SU(2)_1] / SU(2)_3$ and 
$[SU(2)_1 \times SU(2)_2 \times SU(2)_1] / SU(2)_4$. Third, their 
generalisation to higher levels
is proving to be a highly nontrivial problem.
The proof given in \cite{anne} uses infinite product representations of 
level 
$m=3,4,5$ Virasoro characters, as well as the celebrated Jacobi triple
identity and standard properties of generalised theta functions \cite{KP}.
However, it does not shed any light on the underlying structure of these
identities, and therefore offers no clue on how to generalise them to
higher values of the level. \\

In this paper, we use the powerful machinery of number theory \footnote{
A standard textbook discussing the tools needed is \cite{Lang}} to prove
the same identities, and we provide a solid framework within which
more identities can be unveiled.  We establish relations between two
imaginary quadratic extensions over ${\bf{\Qop}}$, which we call $\f$ and
$\fp$ throughout. Given a Galois extension ${\bf N}$ of $\Qop$ which
contains $\f$ and $\fp$, we define two  subgroups of the Galois
group $\Gamma =\gal ({\bf N}/\Qop)$ to be $\De =\gal({\bf N}/\f)$ and
$\Dp =\gal({\bf N}/\fp)$. The deep roots of the Virasoro identities
considered lie in the ability to identify, given ${\bf N}$, pairs of
characters $\chi_{\rm gal}$ and $\chi'_{\rm gal}$ of dimension one on
$\De$ and $\Dp$, which induce the {\em same} character on $\Gamma$,
 \ben
{{\chir}_{\rm gal}} _{\uparrow _{\De} ^{\Gamma}}=
{{\chir'}_{\rm gal}} _{\uparrow _{\Dp} ^{\Gamma}}.
\label{induced}\een
Now, by a standard result, the Artin L-function of an induced
character coincides with the Artin L-function of the original character.
So, given (\ref{induced}), one has,
 \ben L(\chir_{\rm gal})=L({{\chir}_{\rm gal}} _{\uparrow _{\De}
^{\Gamma}})=L({{\chir'}_{\rm gal}} _{\uparrow _{\Dp} ^{\Gamma}})=
L({\chir '}_{\rm gal}).  \label{Lfunct}\een
 As explained later, the L-functions appearing in the expression
(\ref{Lfunct}) are related to ray class theta functions. The latter
obey nontrivial identities which can be derived from the relations
(\ref{Lfunct}) as $\chir_{\rm gal}$ and $\chir'_{\rm gal}$
vary. Theorem 4.3 describes these identities and offers a practical
way to identify relations between the generalised theta functions in
terms of which the Virasoro characters can be expressed.  However its
formulation requires some ground work in number theory.  We introduce
the necessary mathematical jargon and results without the rather
technical proofs, which will be presented elsewhere \cite{Steve}.\\

The paper is organised as follows. Section 2 sets the notations for
coset theta functions \cite{KP}, and emphasizes their r$\hat{\rm o}$le
in the
description of the Virasoro characters identities considered in
\cite{anne}.  Since Theorem 4.3 is a result about ray class theta
functions, we proceed in Section 3 to rewrite coset theta functions as
ray class theta functions. In order to do so, we introduce ray class
groups over $\f$ with conductor $F$. These provide a classification of
ideals in $\Ok$ (the ring of integers over $\f$) which generalises the
classification modulo $n$ of integers prime to $n$. The r$\hat{\rm o}$le 
of the
modulus is played, in imaginary quadratic fields, by the conductor
$F$. The three theorems at the end of Section 4 are the key to
understanding the underlying algebraic structure of the Virasoro
identities (\ref{id1},\ref{id2}), but are also powerful tools in
searching for more such identities. The first two subsections of
Section 4 provide a brief discussion of the ray class characters and
norm maps required in Theorems 4.2, 4.3 and 4.4. We show in Section 5 how 
the Virasoro identities presented in
\cite{anne} arise as a consequence of Theorems 4.3 and 4.4
when $\f =\Qop[\sqrt{-2}]$ (resp. $\f=\Qop[\sqrt{-30}]$) and 
$\fp =\Qop[i]$
(resp. $\fp =\Qop[\sqrt{-10}]$).  As an illustration of
the power of the tools developed in this paper, we also provide 
an infinite family of identities between unitary minimal
Virasoro characters at level $m=3$ and unitary minimal Virasoro
characters at level $m=4a^2$ for $a$ odd and $1+4a^2={a'}^{2}p$ with $p$
prime. The first member of this family, at $a=1$, is the collection of
identities (\ref{id1}).

\section{Coset theta functions}
\renewcommand{\theequation}{2.\arabic{equation}}
\setcounter{equation}{0}
 
The $\frac{m(m-1)}{2}$ independent unitary minimal Virasoro characters at 
level $m$ ($m \ge 2, m \in \Nop$) are analytic functions of the variable
$q=e^{2i \pi \tau}, \tau \in \Cop, {\rm {Im}}\tau \ge 0$ defined by,
 \ben
\chi^{Vir~(m)}_{r,s} (q) = \eta^{-1}(q) \left [
\theta_{r(m+1) -sm,m(m+1)} (q) - \theta_{r(m+1)+sm,m(m+1)}(q) \right],
\label{vir}
\een
with the integers $r$ and $s$ in the following ranges,
$$
r=1,2,\ldots,m-1;~~~~~~~s=1,\ldots,r.
$$
The generalised theta functions at positive integer $k$ are  
given by,
 \ben
\theta_{\l,k} (q) = 
\sum_{n \in \Zop} q^{k\left ( n+\frac{\l}{2k}\right )^2}
\label{theta}
\een 
for $\l$ integer, and the Dedekind eta function $\eta (q)$ can be rewritten
as the difference of two generalised theta functions at level 6,
 \ben
\eta (q)= q^{\frac{1}{24}} \prod_{n=0}^\infty \left (1-q^{n+1}\right)=
\theta _{1,6}(q)-\theta_{5,6}(q).
\label{eta}\een
{}From (\ref{vir}) and (\ref{eta}), it is easy to see that the identities presented in \cite{anne}, namely,
 \ben
\chi^{Vir~(3)}_{1,i} (q) = \epsilon_{ijk} (-1)^{j+k} 
\chi^{Vir~(4)}_{j,4} (q)~\chi^{Vir~(4)}_{k,2} (q)
\label{id1}
\een
and
\begin{eqnarray}
\label{id2}
\chi^{Vir~(4)}_{1,1} (q)\pm \chi^{Vir~(4)}_{3,1} (q) & = & 
\left[\chi^{Vir~(3)}_{1,1}(q)\pm \chi^{Vir~(3)}_{2,1}(q)\right]~
\left[ \chi^{Vir~(5)}_{2,2}(q)\mp\chi^{Vir~(5)}_{3,2}(q)\right], \nn
\chi^{Vir~(4)}_{1,2} (q)\pm \chi^{Vir~(4)}_{3,2} (q) & = & 
\left[\chi^{Vir~(3)}_{1,1}(q)\pm \chi^{Vir~(3)}_{2,1}(q)\right]~
\left[ \chi^{Vir~(5)}_{1,2}(q)\mp\chi^{Vir~(5)}_{4,2}(q)\right], \nn 
\chi^{Vir~(4)}_{2,1} (q) & = & \chi^{Vir~(3)}_{2,2}(q)~
\left[ \chi^{Vir~(5)}_{2,1}(q)-\chi^{Vir~(5)}_{3,1}(q)\right], \nn
\chi^{Vir~(4)}_{2,2} (q) & = & \chi^{Vir~(3)}_{2,2}(q)~
\left[ \chi^{Vir~(5)}_{1,1}(q)-\chi^{Vir~(5)}_{4,1}(q)\right], 
\end{eqnarray}
can be rewritten as identities involving quadratic expressions in 
generalised 
theta functions at different levels. For instance, (\ref{id1}) becomes,
 \ben
V(1,2)V(4-3i,3)=\epsilon_{ijk}(-1)^{j+k} V(5j-16,4)V(5k-8,4)
\label{idp1}\een
where we have introduced the V functions,
 \ben V(r,m)=\theta_{r,m(m+1)}-\theta_{r(2m+1),m(m+1)},
 \label{notation}
 \een
with the properties,
\bea
V(r,m)&=&V(-r,m)=V(r+2km(m+1),m),~~~~k \in \Zop;\nn
V(r,m)&=&-V(r(2m+1),m).
\label{properties}
\eea
(Here, and in future, we will not express the variable $q$.) We note
that,
 \bea 
&& \chi^{Vir(m)}_{r,s}=V(r(m+1)-sm,m)/\eta\quad{\rm and}\quad
 \chi^{Vir(m)}_{r,r}=V(r,m)/\eta ,\nn
&&\eta = \theta_{1,6}-\theta_{5,6}=V(1,2).
 \label{VVir}
 \eea

We first remark that the generalised theta functions (\ref{theta}) are 
coset theta functions in the following sense. 

Let $X$ be a suitably sparse subset of a real or hermitian, positive
definite, inner product space $V$. Choose $d \in \Rop^+$. We define,
 \ben
\theta(X;d)= \sum_{x \in X} q^{\parallel x \parallel^2/d},
\label{cotheta}\een
and we note that for a scalar $\alpha$, we may define 
$\alpha X=\{ \alpha x
\mid x \in X \}$ and then,
 \ben
\theta(\alpha X;\mid \alpha \mid^2 d)=\theta(X;d).
\label{multi}\een
When $X$ is a coset $v+\Lambda $ of a lattice $\Lambda $ in $V$, 
we shall call the function
$\theta(X;d)$ a {\em coset theta function}.  If $\Lambda '$ is a
lattice in another space $V'$, and if $v' \in V'$, we find that 
$\Lambda \times
\Lambda '$ is a lattice in $V \oplus V'$, and that 
$(v+\Lambda) \times
(v' +\Lambda ') = (v,v') +\Lambda \times \Lambda'$. We then get,
 \ben
\theta((v,v')+\Lambda \times \Lambda';d)=\theta(v+\Lambda;d)\theta(v'+\Lambda';d).
\label{product}
\een
{}From (\ref{theta}), (\ref{cotheta}) and (\ref{multi}), we see that,
$$
\theta_{r,k}=\theta(\frac{r}{2k}+\Zop;\frac{1}{k})=
\theta(\frac{r}{2\sqk}+\sqk \Zop;1),
$$
i.e. the generalised theta function $\theta_{r,k}$ is a theta function for
the coset $X=\frac{r}{2k}+\Zop$ of the lattice $\Zop$ in $V=\Rop$. Then
we find, identifying $\Rop \oplus \Rop$ with $\Rop \oplus i\Rop = \Cop$,
\bea
\theta_{r,k}\theta_{s,\l }&\stackrel{(\ref{product})}{=}&
\theta((r/2\sqk )\pm i(s/2 \sql )+(\sqk \Zop+i\sql \Zop);1)\nn
&\stackrel{(\ref{multi})}{=}&
\theta((r\lambda \l_0\pm s\mu\sqD)
+2h\lambda\l_0\mu \langle \mu k_0,\lambda \sqD\rangle _{gp};
4k\l \l_0/h)\nn
&\=&\theta(\a +J;d)
\label{coset}\eea
where $k,\l \in \Nop$, $h=\hcf~(k,\l )$, $\bar{k}=k/h=\mu^2 k_0$ and $\bar{\l}=\l/h=\lambda ^2 \l _0$ with $k_0$ and $\l _0$ square free
as is $D=-k_0\l_0$. So the product of two generalised theta functions
$\theta_{r,k}\theta_{s,\l}$ in $\Rop$ is a theta function for the coset
$X=v+\Lambda \= \a +J$ of a lattice
 \ben
J = 2h\lambda\l _0\mu \langle \mu k_0,\lambda \sqD \rangle _{gp}
\label{lattice}\een
with
$$
\langle \mu k_0,\lambda \sqD \rangle _{gp}=
\{ \mu k_0 n+\lambda \sqD m \mid n,m \in \Zop \}
$$
and
$$
\a = r\lambda \l_0\pm s\mu\sqD .
$$
The next step is to rewrite this coset theta function in terms of a 
ray class
theta function in imaginary quadratic fields. In order to do so, we 
introduce
in the next section the relevant definitions and properties of 
imaginary quadratic fields.

\section{From coset to ray class theta functions}
\renewcommand{\theequation}{3.\arabic{equation}}
\setcounter{equation}{0}

Let $\f$ and $\fk0$ be two fields and $\f \supset \fk0$. Then $\f$ is
called an {\em extension field} of $\fk0$.  The {\em degree of the
extension}, noted $[\f : \fk0 ]$, is the dimension of $\f$ as a vector
space over $\fk0$.  In particular, if $[\f : \fk0 ]=2$, the extension
is a {\em quadratic} extension. In this paper, we work mainly with
imaginary quadratic extensions of $\Qop$ and we introduce the basic
definitions and concepts as applied to them.

\subsection{Ideals and prime factorization in imaginary quadratic fields.}

Let $D$ be a negative integer with no square factor other than 1, and put
 $$
\f = \Qop[\sqD]=\{a+b\sqD \mid a,b \in \Qop\},
$$
 an imaginary quadratic extension of $\Qop$. We define $\Ok$ to be the
ring of all algebraic integers in $\f$. This means that if $D \equiv 2
~{\rm or}~ 3 ({\rm mod} 4)$,
 \ben
\Ok = \Zop [\sqD]=\{a+b\sqD \mid a,b \in \Zop\}=
\langle 1, \sqD \rangle _{gp},
 \label{algint}
 \een
and if $D \equiv 1 ({\rm mod}4)$,
 $$
\Ok = \Zop\left[{\textstyle\frac12}(1+\sqD)\right]
=\left\langle 1, {\textstyle\frac12}(1+\sqD)\right\rangle _{gp}.
$$
It is easy to see that every number in $\f$ is the ratio of two 
algebraic integers. A unit of $\Ok$ is an element of $\Ok$ whose 
reciprocal lies
in $\Ok$. The group of units of $\Ok$ is denoted $\Ok^{\times}$. We note 
in passing,
\begin{Proposition} The units of $\Ok $ are those roots of unity 
which lie in $\f$, vis.
$\{\pm1,\pm i \}$ if $D=-1$, the 6th
roots of unity if $D=-3$ and $\{\pm1 \}$ otherwise.
\end{Proposition}

A {\em fractional ideal I} of $\Ok$ generated by 
$\alpha_1,...,\alpha_n \in \f$ is the set of $\Ok$-linear combinations of 
the generators
$\alpha_i$:
 $$
I = (\alpha_1,...,\alpha_n)_{\Ok} \= 
\{ \sum_{i=1}^n \gamma_i \alpha_i \mid \gamma_i \in \Ok \}.
$$
 In other words, $I$ is a lattice spanning $\f$ which is closed under
multiplication by elements of $\Ok$. A fractional ideal
 $(\alpha)_{\Ok}=\alpha\Ok$, generated by a single
 $\alpha \in \f \setminus \{0\}$ is called a {\em principal ideal}. Note 
that 
 \ben
  \a \Ok = \beta \Ok \iff \a /\beta \in \Ok^\times.
 \label{princequal}
 \een
  We write $\I$ for the set of all fractional ideals of $\Ok$
and $\P$ for the subset of those which are principal.  $\I$ is an
abelian group under ideal multiplication, whose neutral element is
$\Ok$,
  $$
I.J=(\alpha_1,...\alpha_n)_{\Ok}.(\beta_1,...,\beta_m)_{\Ok}=\{ \sum _{i=1}^n\sum_{j=1}^m
\gamma_{ij}\alpha_i \beta_j \mid \gamma_{ij} \in \Ok \}.
$$
 In particular, $\a\Ok.\b\Ok=\a\b\Ok$, so $\P$ is a subgroup of $\I$.
The {\em ideal class group}, ${\cal C}({\bf K})$, is defined to be the
quotient group of $\I$ by $\P$.\\

 If $I\in\I$ is a subset of $\Ok$, then $I$ is said to be an {\em
integral ideal} of $\Ok$. By a {\em prime ideal} of $\Ok$ we shall
mean an integral ideal which is contained in no other apart from
$\Ok$. (Technically, $\{0\}$ is also a prime ideal of $\Ok$ but we
shall ignore this fact.)  Every fractional ideal $I\in\I$ may be
uniquely factorized as a product of integer powers of prime ideals
$Q_i$, $$ I = \prod_{i=1}^m Q_i^{v_{Q_i}(I)}.  $$
 The integer $v_{Q_i}(I)$ is called the {\em valuation} of $I$ at the 
prime
ideal $Q_i$. Thus, only finitely many of the valuations $v_Q(I)$ are
non-zero.
 As one might expect, $I\in\I$ is integral if and only if $v_Q(I) \ge
0$ for all prime ideals $Q$. If $I,\, J\in\I$, we say that $I$ divides
$J$ (we write $I\mid J$) if $JI^{-1}$ is an integral ideal. It follows
 that
 $$
 v_Q(I) \le v_Q(J) ~\forall\, Q \iff I \mid J\iff I \supset J. 
 $$
 Therefore the least common multiple of a pair of ideals $I$ and $J$
is the largest ideal contained in them both and their highest common
factor  is the smallest lattice which contains them both. Thus
 $$
 {\rm lcm}(I,J)=I\cap J\quad {\rm and}\quad
\hcf(I,J) = I+J=\{a+b \mid a \in I, b \in J \}. 
 $$
$I$ is said to be {\it coprime\/} to $J$ if their \hcf\ is $\Ok$. 
Note that in this case
 \ben
 I\cap J={\rm lcm}(I,J)=IJ.
 \label{coprimelcm}
 \een

 A useful parameter of integral ideals is the norm. If $I$ is an
integral ideal of $\Ok$, we define its {\em norm} ${\cal{N}}(I)$ to be
the (finite) number of elements in the quotient group $\Ok/I$. The
norm of a principal ideal $\a\Ok$ is easily found since ${\cal{N}}
(\a\Ok)=| \a |^2$.  Since any fractional ideal $J$ is a quotient
$I{I'}^{-1}$ of integral ideals, we may define ${\cal{N}}(J)={\cal{N}}(I)/{\cal{N}}(I')$. This
definition is unambiguous since, in fact, the norm is multiplicative:
 \ben
{\cal{N}}(IJ)={\cal{N}}(I){\cal{N}}(J).  \label{norm}
 \een
{} From this we see, also, that the norm of an integral ideal gives a
clue as to its prime factors since (in a quadratic field) the norms of
these must be $p$ or $p^2$ for some prime number $p$.

 Indeed, all prime ideals of $\f$ occur as prime factors of $p\Ok$ for
some prime number $p$. Either there is no ideal of norm $p$ and $p\Ok$
is itself prime or there is such an ideal, $P_p$. In this case,
 \ben
 p\Ok=P_p\bar{P}_p,
 \label{primedecomp}
 \een
 and there are exactly one or two ideals of norm $p$ according as
$P_p=\bar{P}_p$ or not. In the rest of this paper $P_p$ will stand for
a prime ideal of $\Ok$ of norm $p$ (if there is one). We will make no
explicit choice unless this is necessary.

If $p$ is an odd prime integer, then the number of ideals of norm $p$
in $\Ok$ is the number of solutions to the congruence $x^2\equiv D$ mod
$p$. (That is, $1+\left( \frac{D}{p} \right)$ where
 $\left(\frac{D}{p} \right)=0$ or $\pm 1$ is the quadratic residue
symbol, cf.  subsection 4.1). (In all the cases that we will examine 
there
will be just one ideal of norm~2.) 

\subsection{Ray class groups}

 Ray class theta functions are based on ray classes which we now
introduce.  These classify ideals in a way which
generalises the classification, modulo $n$, of integers prime to $n$.
For ideals of $\f$, the modulus $n$ is replaced by an integral
ideal $F$, called the {\em conductor}, and we work in
the subgroup $\IF$ of quotients of those ideals which are prime to
$F$, that is,
 $$
 \IF = \{ I \in \I \mid v_P(I) =0 ~{\rm if}~P \mid F\}.  
 $$
 We identify a subgroup $\f_{1,F}$ of $\f^\times$ of elements which
are, in some sense, 1 modulo $F$. These are quotients of elements of
$\Ok$ which are congruent mod $F$ and coprime to  $F$. Thus,
 \ben
   \f_{1,F}= \{ \lambda /\mu
   \mid \lambda - \mu \in F,\, 
     \mu\Ok+F=\Ok, \lambda\mu\ne0 \}.
 \label{PFdef}
 \een
 We put $\PF=\{\a\Ok\mid\a\in\f_{1,F}\}$. Then the {\em ray class
group} of $\Ok$ with conductor $F$ is the quotient group,
 $$
 \CF = \frac{\IF}{\PF}.
$$
  The ray class group is, in fact, a
finite group. Note that, $ {\cal{C}}_{\Ok}(\f)={\cal C}(\f).$

 Each element of $\CF$ is a coset $I\PF $, for some $I \in \IF$. We
denote this coset more compactly as $[I]_F$. We refer to it as the 
{\em ray
class of $I$ with conductor $F$}.

 We shall work, in particular, with $\CPF$, the subgroup of $\CF$ of
ray classes of principal ideals. We shall shorten our notation of such
classes by writing $[\g]_F$ for $[\g\Ok]_F$. Clearly, $\CPF$ is the
kernel of the group homomorphism from $\CF$ to ${\cal{C}}(\f)$ given
by $[I]_F\mapsto[I]_\Ok$. This homomorphism is, in fact, surjective and 
so
we have a short exact sequence of groups,
 \ben
   0\rightarrow\CPF\rightarrow \CF
    \stackrel{{\rm red}_{\Ok}^F}{ \rightarrow}
    {\cal{C}}(\f)\rightarrow 0.
  \label{fullreduction}
  \een \

We conclude with two useful lemmas (with $F$ integral, as above).

\begin{Lemma} Let $H\in \I$.
 (i) If $\a\Ok+FH=H$ then $\a H^{-1}\in\IF$.

 (ii) If, further, $\b-\a\in FH$ then $\b/\a\in\f_{1,F}$
and so $[\b H^{-1}]_F=[\a H^{-1}]_F$.

 (iii)  Conversely, if $\a$ and $\b\in H$ and $\b/\a\in\f_{1,F}$ 
then $\b-\a\in FH$.
 
\end{Lemma} 

{\bf Proof:} (i) $\hcf(\a H^{-1},F)=\a H^{-1}+F=(\a\Ok+FH)H^{-1}= \Ok$.

\medskip \noindent(ii) Since ${\cal C}(\f)$ is finite, $H^n=\g\Ok$ for
some positive integer $n$ and $\g\in\f$.

Put $\lambda=\b\a^{n-1}/\g$ and $\mu=\a^n/\g$. Since $\a\in H$ and
$\b\equiv\a$ modulo $FH$, we have
 $$\lambda\equiv\b\a^{n-1}/\g\equiv\a^n/\g\equiv\mu\ {\rm modulo}\ 
 FH^n/\g=F.$$
Moreover,  $\mu\Ok+F=(\a H^{-1})^n+F=\Ok$. So
 $\b/\a=\lambda/\mu\in\f_{1,F}$.

\medskip\noindent
(iii) Now $\b=\a(\lambda/\mu)\Ok$, where $\lambda-\mu\in F$ and
$\mu\Ok+F=\Ok$. Whence, by (\ref{coprimelcm}), $\mu\Ok\cap F=\mu F$.

But $\b-\a\in H$ and also $\b-\a=(\lambda-\mu)\a/\mu\in FH(1/\mu)$.
 So
 $$\b-\a\in H\cap FH(1/\mu)=(1/\mu)H(\mu\Ok\cap F)=
(1/\mu)H(\mu F)=FH.$$

\begin{Lemma}

 Suppose that
$\a_1$, $\a_2$, $\a_3$ and $\a_4\in\f\setminus FH$ and $\a_i\Ok+FH=H$. 
Then
 
 (i) $\a_i/\a_j=(\a_iH^{-1})(\a_jH^{-1})^{-1}\in\IF$.

\noindent And if $\a_1\a_2\equiv\a_3\a_4$ modulo $FH^2$ then 

 (ii) $[\a_1/\a_4]_F=[\a_3/\a_2]_F$.
 
\end{Lemma}
 
{\bf Proof:} (i) is clear from Lemma 3.2(i). From Lemma 3.2(ii)
 --- with $H^2$ instead of $H$ ---
 $$[\a_1\a_2H^{-2}]_F=[\a_3\a_4H^{-2}]_F$$
 and (ii) follows from this.

\subsection{The ray class theta function.}

{}For any ray class $x$ in the ray class group $\CF$, we define the 
{\em ray class theta function} of $x$ with scale factor $d\in\Rop^+$ as
follows,
 $$ \theta(x;d)=\sum_{I \in x,I \subset \Ok} q^{{\cal {N}}(I)/d}.$$
 Here ${\cal {N}}(I)$ is the norm of $I$ (see (\ref{norm})).
More generally, if $W=\sum_{x \in \CF} n_x x $ is a formal complex
linear combination of ray classes, we write,
 \ben
\theta(W;d)=\sum_{x \in \CF} n_x \theta (x;d).
\label{thetaW}\een
  Also, abusively, if $X$ and $Y$ are subsets of $\CF$, we can regard them 
 as standing for the sums of their elements so that, for instance,
$$
\theta(X-Y;d)=\sum_{x \in X}\theta(x;d)-\sum_{y \in Y}\theta(y;d).
$$

Now the quadratic field coset theta functions as in (\ref{coset}) can be
rewritten as ray class theta functions in the following way. (We do
not tackle here the most general cases as we assume that the lattice $J$ 
of (\ref{lattice}) is an
ideal of $\Ok$. In the general case the coset must first be split up
as a union of cosets of ideals (c.f. \ref{ThetaDecomp}) and a sum of
ray class theta functions will be obtained.)

 Let $J \in\I$ and $\a\in \f\setminus J$.
 Put $H=\hcf(\a\Ok,J)=\a\Ok+J$ and $F=JH^{-1}$, an integral
ideal of $\Ok$.  Thus, by Lemma 3.2(i),
$  [\a H^{-1}]_F\in\CF$

  Let $(\Ok ^{\times})_{F}$ be the group of those
units of $\Ok$ which are 1 modulo $F$ and let  $w_F$ be its order.
We have the following
relation between coset and ray class theta functions,

\begin{Theorem} With $\a$, $J$, $F$, and $H$ as above.
 \ben
\theta(\a +J;d)=w_F\theta([\a H^{-1}]_F;d/{\cal {N}}(H)),
\label{bridge}\een
\end{Theorem}

{\bf Proof:}
 Put $Y$ for the set of all integral ideals in  $[\a H^{-1}]_F$.
 
 The proof of the theorem is effected by the following lemma
which sets up a $w_F$ to 1 correspondence between the terms in the two
theta sums and in which, since ${\cal N}(\b H^{-1})=|\b|^2/{\cal
N}(H)$, the powers of $q$ are scaled by $1/{\cal N}(H)$.

\begin{Lemma} The map
 $f:\a + J \rightarrow Y$ defined by $f(\b)=\b H^{-1}$
is  $w_F$ to 1 and surjective.
\end{Lemma}          

{\bf Proof:} By Lemma 3.2(ii), if $\b\in\a+J$ then $f(\b)\in
[\a H^{-1}]_F$.
 But  $\b\in\a\Ok+J=H$. So $f(\b)\subset\Ok$. Thus
 $f(\b)\in Y$ and $f$ is well-defined.

  Let $I\in Y$. Then $I(\a H^{-1})^{-1}\in\PF$. So,
$IH\a^{-1}=\delta\Ok$, where $\delta\in\f_{1,F}$. Thus $IH=\b\Ok$ with
$\b=\a\delta$. 

Now $\a\in H$ and $\b\in IH\subset H$. So, by Lemma 3.2(iii),
$\b\in\a+FH=\a+J$. Of course, $f(\b)=\b H^{-1}=IHH^{-1}=I$.
 Therefore $f$ is surjective.

 To complete the proof we show that 
 $$f^{-1}(I)=\b(\Ok ^{\times})_{F}.$$
 Suppose $\omega\in(\Ok^\times)_F$. Then $\omega\in1+F$ and so,
 since $\b F\subset HF=J$, $\b\omega\in\b+J=\a+J$. Moreover, by
(\ref{princequal}), $f(\omega\b)=f(\b)=I$.
 Thus $f^{-1}(I)$ contains $\b(\Ok ^{\times})_{F}$.

 We must show the reverse inclusion. If $\b'\in f^{-1}(I)$ then,
certainly, $\b'\Ok=\b\Ok$ so, by (\ref{princequal}), $\b'=\omega\b$
with $\omega\in\Ok^\by$. But, by Lemma 3.2(ii) with $\b'=\a$,
$\omega=\b'/\b\in\f_{1,F}$. Thus, applying Lemma 3.2(iii) with $H=\Ok$,
$\omega\equiv1$ modulo $F$. Thus $\b'\in \b(\Ok ^{\times})_{F}$, as
required.

\subsection{An example of reduction to ray classes}

As an illustration of use of the above identity and of (\ref{coset}),
and as a first step in proving the identities (\ref{id1}), we reduce a
generic theta product occurring in the L.H.S. of (\ref{id1}) to a
coset theta functions and then (in the main case) to a ray class theta
function.

{}First, by (\ref{coset}),
 \ben
\theta_{(1+6a),6}\theta_{s(1+6b),12}=\theta(\a_{s}(a,b) + J;d)
\label{coset1}\een
 where (since, taking $k=6$ and $\l=12$ we have $h=(6,12)=6$,
$\lambda=\mu=k_0=1$ and $D=-2$) $\a \equiv
\a_{s}(a,b)=2(1+6a)+s(1+6b)\sqrt{D}$, $J=24\langle1,\sqrt{D}\rangle$
and $d=96$.

 We note that for $a,b$ taking the values $0$ or $1$, and for $s=1,
-2, -5$, the L.H.S. of (\ref{coset1}) gives all the terms in the three
products $V(1,2)V(4-3i,3), i=1,2,3$, which are the left hand sides of
(\ref{id1}). We recall that generalised theta functions satisfy the
following relations,
 $$
\theta_{\l,k}=\theta_{-\l,k}=\theta_{\l +2k,k}.
 $$
 So we work in the imaginary quadratic field $\f=\Qop (\sqrt{-2})$ and
we introduce the notation $\r = \sqrt{-2}$. Then, from (\ref{algint}),
$\Ok=\langle1,\r\rangle$ and $J=24\Ok$. We find (see
\ref{primedecomp}) prime factorisations $2\Ok=P_2^2$ and
$3\Ok=P_3\bar P_3$, where $P_2=\r\Ok$ and $P_3=(1+\r)_{\Ok}$
with norms 2 and 3, respectively. So $J=P_2^6 P_3\bar
P_3$.

 In order to apply (\ref{bridge}), we first need to determine the
ideal $H$, which is the \hcf\ of $J$ and the principal ideal
generated by $\alpha_{s}(a,b)$. Suppose that $s\equiv 1\mod 3$ and
assume for the moment that $s$ is odd (so $s\equiv1\mod6$).  Then $\a=
2+\r+6\b$ for some $\b\in\Ok$ and $2+\r=\r(1-\r)$. So
 $$
 \a\Ok=P_2\bar P_3(1-\r(1+\r)\b)_\Ok,
 $$
 where the last ideal is clearly
not contained in $P_2$ or $P_3$. Thus
 $H=P_2\bar P_3$ and, using
(\ref{norm}), the norm of $H$ is,
$$
{\cal{N}} (H) = {\cal{N}} (P_2) {\cal{N}} (\bar{P}_3) = 2 \times 3 = 6 .
$$
The conductor (i.e. $F$ of (\ref{bridge})) is
$$
JH^{-1}=P_2^6P_3\bar P_3(P_2 \bar{P}_3)^{-1}=P_2^5P_3=4P_2P_3.
$$
 Thus, by (\ref{bridge}), the coset theta function (\ref{coset1})
reduces, for $s\equiv1\mod6$, to the ray class theta function,
$$
 \theta_{(1+6a),6}\theta_{s(1+6b),12}
 =\theta\left(\left [ \frac{\alpha_{s}(a,b)}{\r (1-\r)}
\right]_{4P_2P_3}; 16\right). 
$$

\subsection{Description of principal ray classes and change of conductor.}

Let $\f$ and $F$ be again as in subsection 3.3. A standard result
(though not difficult to prove) is the following description of the
units of the quotient ring $\Ok/F$:
 $$
 (\Ok/F)^\times=\{\a+F\mid \a\Ok+F=\Ok\}.
 $$ Combining Lemma 3.2 (with $H=\Ok$) with (\ref{princequal}), one can
obtain the following exact sequence  describing $\CPF$,
 \ben
0 \rightarrow (\Ok ^{\times})_{F} \rightarrow \Ok ^{\times}
\rightarrow (\Ok  /F)^{\times} \stackrel{\pi_F}{\rightarrow} 
\CPF \rightarrow 0
\label{princseq}\een
 Here the first map is inclusion, the second takes $u $ to
$u+F$ and $\pi_F$ takes $\a +F$ to $[\a]_F$.\\

In our analysis of section 5, we will need to relate ray class groups
corresponding to two conductors $\tilde{F}$ and $F$ such that
$\tilde{F} \mid F$. This relation is provided by the 
reduction map,
 \ben {\rm red}_{\tilde{F}}^{F}: \CF \rightarrow
{\cal{C}}_{\tilde{F}}(\f) : [I]_{F} \rightarrow [I]_{\tilde{F}}.
\label{reduction}\een
 This is, in fact, a {\em surjective} group homomorphism of which the
projection of (\ref{fullreduction}) is a special case.

Comparing (\ref{fullreduction}) for $F$ and $\tilde F$ it is easy to
see that the kernel of the reduction map ${\rm red}_{\tilde{F}}^{F}$
of (\ref{reduction}) lies in $\CPF$. From (\ref{princseq}) it now
follows that
 $$
 \ker {\rm red}_{\tilde{F}}^{F}
 =\pi_F\{\a+F\in(\Ok/F)^\times\mid \a\equiv1\mod \tilde F\}.
 $$\\

If $F=Q\tilde{F}$, and $Q$ and $\tilde F$
have no common factors, one gets an isomorphism (the Chinese Remainder
Theorem),
  \ben
 (\Ok /F)^{\times}
 \rightarrow (\Ok /Q)^{\times} \times (\Ok /\tilde{F})^{\times}
 :\a + F \rightarrow (\a+Q, \a +\tilde{F}).
 \label{chinrem}
 \een 
 Clearly, isomorphisms of this sort are useful in the description of
$\CPF$ and we generalise the compact notation introduced above for
classes of principal ideals in the following way. If $\g+F$ is the
pre-image of
 $(\a+Q,\b + \tilde{F})$, we write $[\a ,\beta]_F$ for $[\g]_F$.
Note that for $u \in (\Ok )^{\times}$,
 \ben
 [\a,\beta]_F=[u\a,u\beta]_F.  \label{unitrel}
\een
 It is easy to see that the kernel of ${\rm red}^F_{\tilde{F}}$ is
$\{[\a,1]_F \mid \a + Q \in (\Ok/Q)^{\times}\}$. More generally,
if $A \subset {\cal{CP}}_{\tilde{F}}(\f)$, then
 \ben ({\rm red}^F_{\tilde{F}})^{-1}(A) = \{ [\a ,\beta]_{Q\tilde{F}}
\mid \a +Q\in (\Ok/ Q)^{\times}, \b + \tilde{F} \in A\}. 
 \label{liftofsubgroup} \een

 If ${\cal N}(Q)=p$ then $\Zop/p\Zop$ and $\Ok/Q$ have
the same number $p$ of elements. In fact, since $Q\bar Q=p\Ok$ (see
\ref{primedecomp})), $\Zop\cap Q=p\Zop$ and so the map,
 \ben
 \hbox{$\Zop/p\Zop \rightarrow \Ok /P_p ~{\rm by~} 
n+p\Zop \mapsto n+P_p $},
 \label{modpiso}
 \een is a ring isomorphism. Thus in this case the elements of $\CPF$
may be written $[n,\b]_F$ with $n$ an integer mod $p$.

\section{Relations between ray class theta functions of different 
quadratic fields}
\renewcommand{\theequation}{4.\arabic{equation}}
\setcounter{equation}{0}

The identities (\ref {id1}) and (\ref{id2}) are a consequence of 
relations
between ray class theta functions of two different imaginary quadratic
 fields. 
However, to describe these relations, which are stated in Theorem 4.3,
we need to define ray class characters for these two quadratic fields and
their corresponding L-functions. We also use the theory of Artin 
L-functions
in a crucial step (Theorem (4.2)).

\subsection{Ray class characters}

Let $\chir : \CF \rightarrow \Cop ^{\times}$ be a multiplicative 
character
of the ray class group with conductor $F$. We define the {\em conductor}
$F_{\chi}$ of the character $\chir$ to be the biggest of all ideals $I$
dividing $F$ such that the kernel of $\chir$ contains the kernel of the
reduction map (\ref{reduction}) from $\CF$ to $\CI$. In fact, $F_{\chi}$
is the sum of all such ideals. Since the reduction map is surjective,
 $\chir$ defines a character $\chir_{F_{\chi}}$ on 
${\cal{C}}_{F_{\chi}}({\bf{K}})$
defined by the equation,
 $$
\chir([J]_F)=\chir_{\con\chi}([J]_{\con\chi}).
 $$
Thus, for all $I$ divisible by $\con\chi$, $\chir$ also defines a
character $\chir_I$ of $\CI$ by,
 $$
\chir_I([J]_I)\d=\chir_{\con\chi}([I]_{\con\chi}).
 $$
 Note that $\chir_I$ also has conductor $\con\chi$. We refer to the
collection of all such characters $\chir_I$ as ``{\em the ray class
character $\chir$ for ${\bf{K}}$}''.\\

We are now in a position to specialise to the problem at hand. Let us  
introduce two complex quadratic fields ${\bf{K}}=\Qop[\sqD]$ and
${\bf{K'}}=\Qop[\sqDp]$ where $D$ and $D'$ are negative, square free
integers.  We shall also sometimes consider $D''=DD'/(D,D')^2$, which
is square free and positive, along with the quadratic field
${\bf{K''}}=\Qop[\sqrt{D''}]=\Qop[\sqrt{DD'}]$. Here, $(a,b)$ is the 
highest
common factor of the pair of integers $a,b$.

The integer $\Dt$ is defined to be $D$ or $4D$ according to
whether $4$ divides $( D-1)$
or not. The integers $\Ddt$ and $\Dddt$ are defined similarly. 
(These are the
discriminants of ${\bf{K}}$, ${\bf{K'}}$ and ${\bf{K''}}$ over $\Qop$.)

In order to define a particular ray class character $\psi$ for
${\bf{K}}$, we first describe a function $\phi$, which is,
effectively, the Dirichlet character corresponding to the quadratic
field $\f$.

Let $p$ be a prime number and $m$ an integer not divisible by
$p$. Then the quadratic residue symbol, $\left(\frac{m}{p}\right)$ is 
defined to be 1 or
$-1$ according to whether $x^2\equiv m$ has a solution modulo $p$ or not.

{}For positive integers $n$ prime to $2D$ we define $\phi(n)=\pm1$ 
by the following rules:\\

(i) $\phi(n)$ is the quadratic residue symbol
$\left(\frac{D}{n}\right)$ if $n$ is prime.\\

(ii) $\phi(n)=\phi(p_1)\phi(p_2)...\phi(p_r)$ if the
prime factorization of $n$ is $n=p_1p_2...p_r$.\\

\noindent (In fact $\phi(n)$ depends only on the residue of $n$
 modulo $\Dt$).

The function $\phi '$ is defined similarly to $\phi$ (using $D'$
instead of $D$).\\

Let $F=2DD'\Ok$. We define the character $\psi_F$ on $\CF$ by
 $$
 \psi_F([I]_F)\=\phi'({\cal{N}}(I)),
$$
 where $I$ is integral
and ${\cal{N}}(I)$ is the norm of the ideal $I$ --- defined in the
previous section before (\ref{norm}).  Hence we obtain a ray class 
character
$\psi$ for ${\bf{K}}$. (This is the character corresponding the field
extension ${\bf{KK'}}$ of ${\bf{K}}$.) The conductor of this character
is (as shown in \cite{Steve})
 \ben
F_\psi=2^a \Ddt/(\Dt,\Ddt),
\label{cond1}\een
where $a=1$ if all of $\Dt$, $\Ddt$ and $\Dddt$ are even. Otherwise
$a=0$. The ray class character $\psi'$ for ${\bf{K'}}$ is defined 
similarly, and its conductor is,
 \ben
F_{\psi'}=2^a \Dt/(\Ddt,\Dt).
\label{cond2}\een

\subsection{Galois groups and norm maps}

We first identify the elements of the Galois group of the field
${\bf{KK'}}$ over $\Qop$.  In general, if ${\bf{L}}$ is a field 
extension of
the field ${\bf {L}_0}$, the group of automorphisms of ${\bf{L}}$
which leaves every element of ${\bf{L}_0}$ fixed is called the {\em
Galois group of ${\bf{L}}$ over ${\bf{L}_0}$}, and is denoted
$\gal({\bf{L}}/{\bf{L}_0})$. Its order is at most the degree
of the extension. In particular, the biquadratic extension
${\bf{KK'}}=\{a+b\sqD+c\sqrt{D'}+d\sqD \sqrt{D'} | a,\ b,\ c,\
d\in\Qop \}$ has Galois group $\gal({\bf{KK'}}/\Qop)=\{1,\ \delta,\
\delta ',\ \gamma \}$, where
 $$\hbox{
$\delta:\left\{
\matrix{
   \sqD&\mapsto&\sqD\cr
   \sqrt{D'}&\mapsto&-\sqrt{D'}\cr
        }\right.$,\quad
$\delta ':\left\{
 \matrix{
   \sqD&\mapsto&-\sqD\cr
   \sqrt{D'}&\mapsto&\sqrt{D'}
        }\right.$\quad
 and\quad  $\gamma =\delta \delta '$.}
 $$
 Thus $\gamma$ acts as complex conjugation on both ${\bf{K}}$
and ${\bf{K'}}$ and thus gives the non-trivial element of the Galois
group of either field over $\Qop$.  In what follows we shall use
exponential notation for the action of Galois elements. Thus
$x^\delta$ means $x$ is acted on by $\delta$ and
$x^{1-\gamma}=x(x^\gamma)^{-1}$.

\medskip Secondly, if ${\bf {L}} \supset {\bf {L}_0}$ is an extension
of algebraic number fields, then there exist norm homomorphisms, both
denoted by ${\cal{N}}_{{\bf{L}}/{\bf{L}}_0}$, from the group of units
of ${\bf {L}}$ to the group of units of ${\bf{L}_0}$, and from the
group of fractional ideals of ${\bf{L}}$ to the group of fractional
ideals of ${\bf{L}_0}$. Thus,
 $$\matrix{
 {\cal{N}}_{{\bf{L}}/{\bf{L}}_0} :&{\bf {L}}^{\times}& \rightarrow&
{\bf{L}_0}^{\times}& :
 \lambda &\mapsto &{\cal{N}}_{{\bf{L}}/{\bf{L}}_0} (\lambda),\cr
 {\cal{N}}_{{\bf{L}}/{\bf{L}}_0} :&{\cal{I}}({\bf{L}})&
\rightarrow& {\cal{I}}({\bf{L}}_0)&
 : I& \mapsto &{\cal{N}}_{{\bf{L}}/{\bf{L}}_0}(I).\cr
          }$$
 These norms are related by the fact that they `agree' on elements
$\lambda$ of {\bf L}. That is,
 $$
{\cal{N}}_{{\bf{L}}/{\bf{L}_0}}(\lambda {\cal{O}}_{{\bf{L}}})=
{\cal{N}}_{{\bf{L}}/{\bf{L}_0}} (\lambda) {\cal{O}}_{{\bf{L}_0}}.
$$
Moreover, if $I \in {\cal{I}}({\bf{L}})$, one has,
 $$
{\cal{N}}(I) \Zop = {\cal{N}}_{{\bf {L}}/\Qop}(I).
 $$
If the extension ${\bf{L}}$ over ${\bf{L}_0}$ is quadratic with 
$\gal({\bf{L}}/{\bf{L}_0})=\{1,\ \sigma \}$, then,
for $\lambda \in {\bf{L}}^{\times}$,
  $$
{\cal{N}}_{{\bf{L}}/{\bf{L}}_0}(\lambda)=\lambda ^{1+\sigma}
  $$
 and for $I \in  {\cal{I}}({\bf{L}})$, we have
  $$
{\cal{N}}_{{\bf{L}}/{\bf{L}}_0}(I){\cal{O}}_{{\bf{L}}}=I^{1+\sigma}.
  $$
 This equation determines the norm map on ideals since the map,
 $$
{\cal{I}}({\bf{L}}_0) \rightarrow {\cal{I}}({\bf{L}}) : 
J\mapsto J{\cal{O}}_{{\bf{L}}},
 $$
is injective.
In particular, if ${\bf{L}}={\bf{KK'}}$ and ${\bf {L}_0}={\bf{K}}$, 
one has that,
for $I\in {\cal{I}}({\bf{KK'}})$ and $J\in {\cal{I}}({\bf{K}})$,
 $$
{\cal{N}}_{{\bf{KK'}}/{\bf{K}}}(I){\cal{O}}_{{\bf{KK'}}}=
I^{1+\delta}\quad{\rm and}\quad{\cal{N}}(J)\Ok=J\overline J.
 $$

{}Finally, choose a fractional ideal $F$ in ${\cal{I}}(\f)$ which is
contained in the conductor $F_{\psi}$ (\ref{cond1}), and which is
self-conjugate, i.e.,
 $$
F \in {\cal{I}}({\bf{K}}), \quad F \subset F_{\psi}
 \quad{\rm and}\quad F^{\gamma}\ ( = \overline{F})\  = F.
 $$
We define a subgroup $A_F$ and a coset $S_F$ in the ray class group 
$\CF$ by,
 \ben
A_F \= {\rm{ker}}(\psi_F)^{1-\gamma}=\{ [I/{\overline I}^{-1}]_F \mid  \psi_F([I]_F)=1\},\quad
S_F\=\{[I{\overline I}^{-1}]_F\mid \psi_F([I]_F)=-1\}.
\label{cos}\een
 For
$F'\subset  F_{\psi'}$ we define $A'_{F'}$ and $S'_{F'}\subset 
{\cal{C}}_{F'}
({\bf{K}}')$, similarly.

 Note that if $\psi_F([I]_F)=-1$ then $S_F=(I/\overline I)A_F$. Also
if $F_1$ is a self-conjugate ideal contained in $F$, we find that,
 \ben 
 \hbox{ $A_{F}={\rm red}^{F_1}_F(A_{F_1})$
and $S_F={\rm red}^{F_1}_F(S_{F_1})$.}
 \label{redskews}
 \een

\subsection{The three crucial theorems}

To provide a link between the arithmetic of $\bf K$ and that of $\bf
K'$ we need to give a correspondence between conductors in $\Ok$ and
in $\Okp$.

\begin{Definition}
  We say that a pair $(F,F')$ of self-conjugate ideals, 
$F\in{\cal I}(\f)$
and $F'\in{\cal I}(\fp)$ is {\em admissible\/} if
$F\subset F_\psi$, $F'\subset F'_{\psi'}$ and
 ${\cal{N}}(F)\tilde D={\cal{N}}(F')\tilde D'$.
  \end{Definition}

In this subsection we present our main theorem for producing
identities, Theorem 4.3, which is a practical theorem relating ray
class theta functions of different quadratic fields.  Given an
admissible pair $(F,F')$, the theorem describes coincidences between
certain combinations of ray class theta functions of $\f$ with
conductor $F$ and similar combinations of ray class theta functions of
$\fp$ with conductor $F'$.\\

The origin of these coincidences can be described in terms of 
$L$-functions of
 characters for the ray class groups $\CF$ and
${\cal{C}}_{F'}({\bf{K}}')$. We first define such $L$-functions, then 
give a crucial relation between them in Theorem (4.2).

Given the definition (\ref{thetaW}) of theta functions for a formal
complex linear combination of ray classes $\theta(W;d)$, one obtains
the corresponding $L$-function $L(W)$ by using the modified Mellin
transform, $M_d$, which sends $q^t$ to $(td)^{-s}$,
 \ben
M_d(\theta(W;d))=L(W)=\sum_{x \in X}n_x L(x),
\label{mellin}\een  
 where $L(x)=\sum_{I \in x, I \subseteq \Ok} {\cal{N}}(I)^{-s}$.

Now, if $\chir$ is a ray class character of $\f$ with conductor
$F_{\chi}$ dividing $F$, we may define the {\em L-function of the ray
class character $\chir$ with conductor  $F$} as follows,
 $$
L_F(\chir) \= \sum_{x \in \CF} \chir(x) L(x) = 
\sum_{I \in \IF, I \subseteq \Ok}\chir([I]_F) {\cal{N}}(I)^{-s}.
 $$
We put $L(\chir)=L_{F_{\chi}}(\chir)$.
We now have,
\begin{Theorem}
 Let $(F,F')$ be an admissible pair and let $\chir$ and
$\chir'$ be characters of $\CF$ and ${\cal{C}}_{F'}({\bf{K}}')$ such 
that

(i) $\chir$ is 1 on $A_F$ and $-1$ on $S_F$.

(ii) For all $I\in {\cal{I}}_{FF'{\cal{O}}_{{\bf{KK'}}}}({\bf{KK'}})$,
$\chir([{\cal{N}}_{{\bf{KK'/K}}}(I)]_F)
=\chir'([{\cal{N}}_{{\bf{KK'/K'}}}(I)]_{F'})$.

Then $L_F(\chir)=L_{F'}(\chir')$.
\end{Theorem}

The proof relies on the theory of Artin $L$-functions. In fact, for a
suitable extension ${\bf{N}}$ of ${\bf{KK'}}$, the characters
$\chir$ and $\chir'$ define, under the Artin correspondence of class
field theory, characters
 $\chir_{\rm gal}$ of $\Delta={\rm Gal}({\bf{N/K}})$ and
 $\chir'_{\rm gal}$ of $\Delta'={\rm Gal}({\bf{N/K'}})$. Now the Artin
$L$-functions of $\chir_{\rm gal}$ and $\chir'_{\rm gal}$ are, by
definition, the $L$-functions of $\chir$ and $\chir'$,
 $$
L(\chir_{\rm gal})=L(\chir),~~~~~L(\chir'_{\rm gal})=L(\chir').
 $$
One proves \cite{Steve} that, under the
conditions of the theorem, $\chir_{\rm gal}$ and $\chir'_{\rm gal}$
induce the same character of $\Gamma=\gal({\bf{N/\Qop}})$,
 $$
{{\chir}_{\rm gal}} _{\uparrow _{\De} ^{\Gamma}}=
{{\chir'}_{\rm gal}} _{\uparrow _{\Dp} ^{\Gamma}}.
$$
Since their Artin
$L$-functions coincide with that of the induced character, one must
have,
 \ben
L(\chir)=L(\chir _{{\rm gal}})=L(\chir '_{{\rm gal}})=L(\chir ').
 \label{Lrel}
 \een
The result follows.\\

We are now ready to state the main theorem of this paper. It is
obtained by summing (multiplied by suitable roots of unity) all the
instances of (\ref{Lrel}) for a given admissible pair of conductors (a
sort of finite Fourier inversion) and applying the inverse of the
Mellin transform given in (\ref{mellin}).

\begin{Theorem}
Let $(F,F')$ be admissible and let $I$ be an integral ideal of 
${\cal{O}}_{\f \fp}$ having no common factor with 
$FF' {\cal{O}}_{\f \fp}$.
Put $J={\cal{N}}_{{\bf{KK'/K}}}(I)$ and 
$J'={\cal{N}}_{{\bf{KK'/K'}}}(I)$.
 Then
  \ben
\theta(A_F[J]_F;d)-\theta(S_F[J]_F;d)=
\theta(A'_{F'}[J']_{F'};d)-\theta(S'_{F'}[J']_{F'};d),
\een
for $d \in \Rop$.
\end{Theorem}

Unfortunately, as will be exemplified in the next section when we
discuss the first set of Virasoro character identities (\ref{id1}),
coset theta functions may give rise, by application of
(\ref{bridge}), to theta functions of ray classes with respect to
non-self-conjugate conductors. The following theorem describes
situations where such theta functions may be replaced by theta
functions of ray classes with self-conjugate conductors. It relies on
the cancellation available from the relationship (for
self-conjugate~$F$) $$ L([\,\overline I\,]_F)=L([I]_F).\nn $$

\begin{Theorem} 
  Let $F$, $P$ and $J$ be (integral) ideals of $\Ok$. Suppose that 
$F$ is
  self conjugate, that $P$ is maximal and prime to $F$ and that $J$ is
  prime to $PF$. 

  Let $B$ be a subgroup of $\CF$ containing both
  $[P/\overline{P}]_F^2$ and $[J/\overline{J}]_F$.  Put $T$ for the
  coset $B[P/\overline{P}]_F$ and put $\tilde{B}$ and $\tilde{T}$ for
  the inverse images of $B$ and $T$ in ${\cal{C}}_{FP}(\f)$.

Then
  \ben
\theta (\tilde{B}[J]_{FP};d)-\theta (\tilde{T}[J]_{FP};d)=
\theta (B[J]_F;d)-\theta (T[J]_F;d).
\een
\end{Theorem}

We shall apply this result in cases where $B=A_F$ and $T=S_F$. We note
that in this case, the conditions of the second paragraph of the 
above theorem hold, provided
$f_\psi\mid F$, $\psi(J)=1$ and $\psi(P)=-1$.

The proofs of the above theorems (or rather the corresponding theorems 
for $L$-functions) may be found in
\cite{Steve}. We now use them to prove the Virasoro identities
(\ref{id1}, \ref{id2}), but also to uncover a whole family of
identities between Virasoro characters at higher levels, of which the
identities (\ref{id1}) are the simplest example.

\section{Virasoro identities}
\renewcommand{\theequation}{5.\arabic{equation}}
\setcounter{equation}{0}

In the next three subsections, we use the algebraic tools developed in
this paper to prove the identities (\ref{id1}). We first show in
subsection 5.1. how to relate ray class theta functions associated
with the two relevant quadratic fields $\f = \Qop[\sqrt{-2}]$ and $\fp
=\Qop [\sqrt{-1}]$ using Theorem 4.3.  We then rewrite the
identities as in (\ref{idp1}). To make contact with Theorem 4.3, we
express the left hand side (resp. the right hand side ) of
(\ref{idp1}) in terms of differences of ray class theta functions for
$\CF$ (resp. ${\cal{C}}_{F'}({\bf{K}'})$) for self-conjugate
conductors $F$ (resp. $F'$). This is carried out in detail in
subsections 5.2. and 5.3.

 Remarkably, the three distinct differences of ray class theta
functions appearing in the RHS of (\ref{idp1}) are obtained each
time one considers quadratic expressions in Virasoro unitary minimal
characters at level $m=4a^2$ for $a$ odd and $1+4a^2=a'^2p$ with $p$
prime, in a way described in Theorem (5.1), subsection 5.3. This does
then provide an infinite class of new identities between the Virasoro
characters at level $m=3$ and Virasoro characters at level $m=36$,
100, $196,\ldots$ We also remark (\ref{lhsexp}) how the LHS of these
identities may be rewritten as sums of Virasoro characters at any of an
infinite series of higher levels $m=675$, 131043$,\ldots$

 The subsection 5.4. gives a somewhat terser account of how to prove
the second set of identities (\ref{id2}). There, the two relevant
quadratic fields are $\f = \Qop[\sqrt{-30}]$ and $\fp =\Qop
[\sqrt{-10}]$.

We recall (see Section 3) that in
what follows, $P_p$ (resp. $P'_p$) stands for a
maximal ideal of $\Ok$ (resp. of $\Okp$) of norm $p$ when $p$ is prime.
 
\subsection{Relations between ray class theta functions}

As already remarked at the end of Section 3, the expression on the
left hand side of the identities (\ref{idp1}) is associated with the
quadratic field $\Qop[\sqD]$ with $D=-2$, while the right hand side is
associated with the quadratic field $\Qop[\sqDp]$ with $D'=-1$. So, in
the notations adopted in this paper, one has $\f =\Qop[\r]$ with $\r
=\sqrt{-2}$, $\Ok=\langle 1, \r \rangle _{gp}$, and the units
of $\Ok$ are $\Ok ^{\times}= \{ \pm 1 \}$. Also, $\fp = \Qop [i]$,
$\Okp = \langle 1,i\rangle _{gp}$ and $\Okp ^{\times}= 
\{ \pm 1, \pm i \}$.

Consider the two ray class characters $\psi $ and $\psi '$ defined in
subsection 4.1. Their conductors are given by the expressions
(\ref{cond1},\ref{cond2}) with $\tilde{D}=-8$, $\tilde{D'}=-4$ (and
$\widetilde {D''}=8$, so $a=1$). Thus,
$$ \eqalign{
   F_{\psi} &=\left(\frac{2[8,4]}{8}\right)_{\Ok}= (2)_{\Ok}\cr
   F_{\psi'}&= \left(\frac{2[4,8]}{4}\right)_{\Okp}=(4)_{\Okp}.\cr}
 $$
 (Note that, especially in subscripts, we shall write principal ideals
$(\a)$ instead of $\a\Ok$ or $\a\Okp)$.

 Both $\Ok$ and $\Okp$ are principal ideal domains, i.e. all their
ideals are principal, so that, using (\ref{princseq}), we can easily
identify the ray class groups of $\f$ and $\fp$ associated with the
above conductors,
$$ \begin{array}{c}
 {\cal{C}}_{F_{\psi}}(\f) =
{\cal{C}}_{(2)}(\f)={\cal{CP}}_{(2)}(\f) \simeq (\Ok /(2))^{\times} =
\langle [1+\r]_{(2)} \rangle _{gp}\\
 {\cal{C}}_{F_{\psi '}}(\fp)
={\cal{C}}_{(4)}(\fp)={\cal{CP}}_{(4)}(\fp) \simeq (\Okp
/(4))^{\times}/\{\pm 1,\pm i \} = \langle [1+2i]_{(4)} \rangle
_{gp}.  
\end{array}$$
 Both groups are of order 2, and so $\psi $ and $\psi '$ are
the only non trivial (1-dimensional) characters of
${\cal{C}}_{F_{\psi}}(\f)$ and ${\cal{C}}_{F_{\psi '}}(\fp)$. We thus
have,
 \bea
 \psi ([\a]_{(2)}) &=& 1~~~{\rm if}~\a \equiv 1~~{\rm
  mod}~(2)\nn &=& -1~~~{\rm otherwise}
\label{char1}\eea
\bea 
\psi '([\a]_{(4)})&=& 1~~~{\rm if}~\a \equiv \pm1\ {\rm or}\ \pm i
  ~~{\rm mod}~(4)\nn
                   &=& -1~~~{\rm otherwise}.
\label{char2}\eea

The next step is to calculate the $A$'s and $S$'s of (\ref{cos}) for
the admissible pairs of ideals $(F,F')$ (see Definition 4.1) that we
shall be using.  Put $P_2=(\r)_{\Ok}$ as before and
$P_2'=(1+i)_{\Okp}$, the prime ideal of norm 2 in $\Okp$ (noting that
${P'_2}^2=2\Okp$). Then $(4P_2,(8)_{\Okp})$ and $((4)_{\Ok},4P_2')$ are
admissible pairs for $\f$ and $\fp$. We concentrate on the first pair
as the data for the second pair will come easily by
(\ref{redskews}). By (\ref{char1}),
 $$
 \ker~\psi_{4P_2} = \{ [ 1+2\a]_{4P_2} \mid \a \in \Ok \}=
 \{[1+2a+2b\r]_{4P_2} \mid a,b \in \Zop \}.
 $$
 Thus, in this case, the $A$-group is trivial,
 \ben
 A_{4P_2}=(\ker\psi_{4P_2})^{1-\gamma }=\{[1]_{4P_2}\},
 \label{Atriv}
 \een
 since $\overline{1+2a+2b\r}=1+2a-2b\r\equiv 1+2a+2b\r \mod
4P_2\,(=(4\r))$ (and $\gamma $ acts as conjugation).  In order to
obtain the coset $S_{4P_2}$ (Definition (\ref{cos})), we must choose
an ideal $I$ in ${\cal{I}}_{4P_2}(\f)$ such that
$\psi\,[I]_{4P_2}=-1$. We take $I=(1+\r)_{\Ok}$. Then
we find that $S_{4P_2}$ consists of the class
 \ben
[I/\bar{I}]_{4P_2} = [(1+\r)(1-\r)^{-1}]_{4P_2}
 =[3(-1+2\r)/9]_{4P_2}= [-3+2\r]_{4P_2}= [5\pm2\r]_{4P_2}.
 \label{Scalc}
\een

We also need $A'_{(8)}$ and its coset $S'_{(8)}$ for $\fp$. Note that now,
 \ben
\ker~\psi'_{(8)} = \{ [1+4\a]_{(8)} \mid \a \in \Okp \}=
\{ [1+4a+4bi]_{(8)} \mid a,b \in \Zop \},
\label{ker}\een
and so 
 \ben
 A'_{(8)}=\ker~\psi'_{(8)})^{1-\gamma}=\{[1]_{(8)}\},
 \label{Ad8}
 \een
 since, modulo $(8)$, 
 $1+4a+4bi\equiv 1+4a-4bi =  \overline{1+4a+4bi}$.
 Moreover, with $I=(1+2i)_{\Okp}$, $\psi'[I]_{(8)}=-1$, and so we find
that $S'_{(8)} $ consists of the class
 \ben
 [I/{\bar{I}}]_{(8)}=[(1+2i)(1-2i)^{-1}]_{(8)}
  =[5 (1+2i)^2/25]_{(8)}=[-1+4i]_{(8)}
  =[1+4i]_{(8)}.
\label{skew1}\een

{}For the second pair of conductors we find that the ideal $(4)_{\Ok}$
divides the ideal $4P_2$, and $4P'_2$ divides $(8)_{\Okp}$. So we can
obtain the new $A$'s and $S$'s by reduction using (\ref{redskews}) $$
A_{(4)}= \{ [1]_{(4)} \}~~~{\rm and}~~~S_{(4)}=[1+2\r]_{(4)}, $$
 \ben
A'_{4P_2'}=(\ker~\psi'_{4P'_2})^{1-\gamma} = \{ [1]_{4P_2'} \}~~~
{\rm and}~~~
S'_{4P'_2}=[5]_{4P'_2}=[3]_{4P'_2}.
\label{skew2}\een

Now take $I$, $F$ and $F'$ in Theorem 4.3 to be successively
 $$\matrix{
  {\cal O}_{\bf KK'},&4P_2&\rm and &8{\Okp};\cr
  (1-\r)_{{\cal O}_{\bf KK'}},&4P_2&\rm and&8{\Okp};&\rm and\cr
  {\cal O}_{\bf KK'},&4\Ok,&\rm and&4P'_2.\cr}
 $$
 This gives the following relations between ray class theta
functions of the fields $\Qop[\r]$ and $\Qop[i]$, 
\bea
\theta([1]_{4P_2};16)-\theta (S_{4P_2};16)&=& \theta
([1]_{(8)};16)-\theta(S'_{(8)};16)\nn
\theta([1+2\r]_{4P_2};16)-\theta ([1+2\r]_{4P_2}S_{4P_2};16)&=&
\theta ([3]_{(8)};16)-\theta ([3]_{(8)}S'_{(8)};16)\nn \theta
([1]_{(4)};8)-\theta (S_{(4)};8)&=& \theta
([1]_{4P'_2};8)-\theta (S'_{4P'_2};8).
\label{relations}\eea

Here, for the second line we have calculated the norms of 
$(1-\r)$ from $\f \fp$ to $\f$ and from $\f \fp$ to $\fp$ as follows
 $$ \eqalign{
{\cal{N}}_{\f \fp/\f}(1-\r)&=(1-\r)^2=-(1+2\r),\cr
 {\cal{N}}_{\f \fp/\fp}(1-\r)&=(1-\r)(1+\r)= 3. \cr}
 $$

\subsection{Reduction of V products to ray class theta functions}

We have shown at the end of section 3 how to rewrite a product of two
generalised theta functions as a ray class theta function, namely,
 \ben
 \theta_{(1+6a),6}\theta_{s(1+6b),12}
=\theta\left(\left [\frac{\alpha_{s}(a,b)}{\r(1-\r)}\right]_{P_3F}; 
16\right),
 \label{coset2}
 \een
with $s\equiv 1\mod 6$ and $F=4P_2$.
 From (\ref{notation}) and  the above relation,
we obtain the V product $V(1,2)V(s,3)$ as a linear combination of 
ray class theta functions,
 \ben
 V(1,2)V(s,3)=(\th_{1,6}-\th_{7,6})(\th_{s,12}-\th_{7s,12})=
\theta(W;16),
\label{firststep}
\een
where
 \ben
W=[\b_s]_{P_3F}(q(0,0)+q(1,1)-q(1,0)-q(0,1))
\in {\cal C}_{P_3F}(\f),
\label{firststepsum}
 \een
 with $\b_s=\a_s(0,0)/((1-\r)\r)$ and 
 $q(a,b)=[\alpha_{s}(a,b)/\alpha_{s}(0,0)]_{P_3F}$.

We express the elements $[\delta]_{P_3F}$ of
${\cal{C}}_{P_3F}(\f)={\cal{CP}}_{P_3F}(\f)$ in the form
$[\a,\b]_{P_3F}$ as explained after (\ref{chinrem}), following the
decomposition
 $$ (\Ok/P_3F)^\times
 \cong(\Ok/P_3)^\times\times (\Ok/F)^\times.
 $$

In order to apply Theorem 4.3 we shall combine the ray classes of
(\ref{firststep}) to make classes with respect to the self-conjugate
conductor $F$ using Theorem 4.4. We take $P$, $B$ and $T$ of that
theorem to be, respectively, $P_3$, $A_{F}\,(=\{[1]_{F}\}$ by
\ref{Atriv}) and $S_{F}\,(=\{[5-2\r]_{F}\}$ by \ref{Scalc}).  The
choice of $P=(1+\r)$ and $T$ are consistent since from (\ref{char1}),
$\psi([1+\r])=-1 $. We first identify $\tilde B$ and $\tilde T$ and
then compare them to the classes $q(a,b)$.

We know that ${\cal N}(P_3)=3$.
 So $(\Ok/P_3)^\times = \{\pm1+P_3\}$ and 
 by, (\ref{liftofsubgroup}),
 $\tilde B$ and $\tilde T$ of Theorem 4.4 are $\{[\pm1,1]_{P_3F}\}$
  and $\tilde B[5-2\r]_{P_3F}$, respectively.

Now $\a_s(1,1)=7\a_s(0,0)$. So
 \ben
q(1,1)=[7]_{P_3F}=[1,-1]_{P_3F}=[-1,1]_{P_3F},
 \een
 by (\ref{unitrel}).
Also, $7\a_s(1,0)=98+7s\r\equiv\a_s(0,1)\mod 24$. So
 $$
 \a_s(0,1)/(\r(1-\r))\equiv7\a_s(1,0)/(\r(1-\r))\mod P_3F
 $$
  and hence
 $$
 q(0,1)=[7]_{P_3F}q(1,0)=[1,-1]_{P_3F}q(1,0).
 $$
Again, $\a_s(0,0)(5-2\r)\equiv\cdots\equiv\a_s(1,0)\mod 24$.
So, in the same way, 
 $$
  q(1,0)=[5-2\r]_{P_3F}q(0,0).
 $$
 Thus
  $$
 \{q(0,0), q(1,1) \} = \tilde B \quad {\rm and}\quad
 \{q(1,0),q(0,1)\} =  \tilde T
  $$
 Putting this information into (\ref{firststepsum}) we get, from
\ref{firststep},
 \ben
V(1,2)V(|s|,3)=\th([\b_s]_{P_3F}\tilde B;16)-
\th([\b_s]_{P_3F}\tilde T;16)
 = \th ([\b_s]_{F};16)-\th([\b_s]_{F}S_{F};16)
\label{mellin1}\een
by Theorem 4.4. Moreover, $\b_s=1$ if $s=1$ and $\b_s=1+2\r$ if $s=5$.

If $s=-2$  one can divide the coset in (\ref{coset1}) by $-\r$ to get,
  $$
 \theta_{(1+6a),6}\theta_{2(1+6b),12}
 =\theta(2(1+6b) - (1+6a)\r +(J/P_2);8). 
 $$
 Here we have the same situation as before with $s_{{\rm new}}=1$ 
except that
the r$\hat{\rm o}$oles of $a$ and $b$ are reversed, the ray class 
conductor is
$4P_3$ and the scale factor $8$ instead of 16. The same analysis goes
through with $F=4\Ok$ and we obtain
 \ben
V(1,2)V(2,3)= \th([1]_{(4)};8)-\th(S_{(4)};8).
\label{mellin2}\een

Thus we have shown that the left hand sides of the identities
(\ref{idp1}) may be rewritten as the left hand sides of the identities
(\ref{relations}).  To complete our proof of (\ref{idp1}) we shall, 
in a
similar manner, show that the right hand sides of (\ref{idp1}), which
may be described by the V products
 \ben
 V(r,4)V(rf,4)+V(7r,4)V(7rf,4)
 \label{RHSrewrite}
 \een
 for $(r,f)=(1,2)$, $(3,2)$ and (1,3), are equal to the right hand
sides of the identities (\ref{relations}).

\subsection{An infinite family of identities}

 In fact the expressions (\ref{RHSrewrite}) are only the first set of
an infinite family of quadratic expressions in the $V$ functions which
reduce to the right hand sides of (\ref{relations}). We now prove this
reduction for the whole family and thus obtain (by \ref{VVir}) an
infinite family of identities between Virasoro characters at level
$m=3$ and products of those at levels $m=4a^2$ where $a$ is odd and
$1+4a^2=p{a'}^2$ with $p$ prime.

\begin{Theorem}
 Let $a$, $a'$ and $p$ be integers such that $a\equiv1\mod 4$, $p$ is
prime and $4a^2+1=pa'^2$. Put $m=4a^2$ and $c=aa'$.  Then, for $r$ odd
and prime to $p$ and $\epsilon=0$ or $1$,
  \ben
\sum_{u=1}^{\frac{p-1}{2}}\sum_{v=0}^{c-1}\sum_{w=0}^{c-1}
 V(c\uh (r+8vp),m)V(c\uh ((2\ah-\e p)r+8wp),m)
 =\th([r]_{F'}-[r\delta]_{F'},2^{4-\e}),
\label{dagger}
\een
 where $\uh =(u+5p(1-u))$ and the ray classes on the right are defined
in $\fp=\Qop[i]$ with 
 $F'={P'_2}^{6-\e}$, $P_2'= (1+i)_{\Okp}$ (as in subsection 5.1) and
$\delta =1+4i$ so that $\{[\delta]_{F'}\}=S'_{F'}$ (see \ref{skew1} and
\ref{skew2}).
 \end{Theorem}
 
Actually, the theorem gives just three different relations. These
correspond to the choices $(r,\epsilon)=(1,0)$, $(3,0)$ and $(1,1)$
and have as right hand sides the three right hand sides of
(\ref{relations}). Moreover if we take $a=1$ and $p=5$, so $m=4$,
$a'=c=1$ and $\hat2=-23$ and put $f=|2a-\e p| = 2$ or 3, then the LHS
in Theorem 5.1 becomes,
 $$
 V(r,4)V(rf,4)+V(23r,4)V(23rf,4).
 $$
 Now we may make this expression more economical by replacing 23 by 7
(since $23\times9=207\equiv7$ modulo $2m(m+1)=40$, $9=2m+1$ and
$V(r(2m+1),m)=-V(r,m)$). So we have,
 \ben
 V(r,4)V(rf,4)+V(7r,4)V(7rf,4)
 =\th([r]_{\tilde{F}'};2^{4-\e})-\th([r]_{\tilde{F}'}S'_{\tilde{F}'};
2^{4-\e}).
 \label{VV}
 \een
 Taking $(r,f)=(1,2),(3,2)$ and $(1,3)$, and using (\ref{mellin1}),
(\ref{mellin2}) and (\ref{relations}) we get the identities 
(\ref{idp1}).\\

Before we can prove Theorem 5.1, we need some preparation.

Suppose that $\Lambda \supset\Lambda '$ are lattices in the inner
product space $V$. Then there is a subset $T\subset\Lambda $ (a
transversal of $\Lambda$ over $\Lambda '$) of the same size as the
quotient group $\Lambda /\Lambda '$ such that $(w+\Lambda ')$ and
$(w'+\Lambda ')$ are disjoint for distinct $w$ and $w'$ in $T$.
Then, for any $v\in V$,
 $$ 
v+\Lambda =\bigcup_{w\in T}(v+w)+\Lambda '
 $$
and so,
 \ben
\theta(v+\Lambda ;d)=\sum_{w\in T}\theta((v+w)+\Lambda ';d).
\label{ThetaDecomp}\een

\begin{Lemma}
Suppose that $k=c^2k'$ with $c$ and $k'\in\Nop$. Choose
$b\in\Zop$ to have no common factor with $c$. Then
 \ben
\sum_{j=0}^{c-1}\th_{cb(r+2jk'),k}=\th_{br,k'}.
\label{ThetaCons}\een
\end{Lemma}

{\bf Proof}: Now
 $$\th_{cb(r+2jk'),k}
 =\th\left({cb(r+2jk')\over2k}+\Zop;\,{1\over k}\right)
 =\th\left(\left({br\over2k'}+bj\right)+c\Zop;\,{1\over k'}\right).$$

Whereas 
 $$\th_{br,k'}=\th\left({br\over2k'}+\Zop;{1\over k'}\right).$$
 But since $b$ is invertible mod $c$, the set
 $\{0$, $b,\ldots ,(c-1)b\}$ is a transversal for $\Zop$ over
$c\Zop$. So the identity follows by (\ref{ThetaDecomp}). As an
immediate consequence we have,

\begin{Corollary}
 If $m(m+1)=k$ above then
  \ben
\sum_{j=0}^{c-1}V(cb(r+2jk'),m)=\th_{br,k'}-\th_{br(2m+1),k'}.
\label{VirCons}\een
\end{Corollary}

We remark, by the way, that taking $m=242$, $c=99$, $k'=6$ and
$b=r=1$ gives a RHS in (\ref{VirCons}) of
$\th_{1,6}-\th_{5,6}=\eta$ and we obtain,
  \ben
\sum_{j=0}^{98}\chir^{Vir(242)}_{99(1+12j),99(1+12j)}=1.
 \een
 This is the first in an infinite series of such identities (though,
presumably not the first with RHS~1). The next, however has $m=23762$ 
and $c=109\times 89=9701$.\\

Again, we may solve the Pellian equation
 $$ (2m+1)^2-48c^2=1 \quad({\rm i.e.}\quad m(m+1)=12c^2)$$
 and choose one of the (infinitely many) solutions such that
$2m+1\equiv 7$ modulo 24. (e.g. $m=675$, $c=175$; $m=131043$,
$c=37829$.) In Lemma 5.2, we now have $k'=12$ and
$b=1$ gives a RHS in (\ref{VirCons}) of
$\th_{r,12}-\th_{7r,12}=V(r,3)$ and we obtain, for instance,
  \ben
\sum_{j=0}^{174}\chir^{Vir(675)}_{175(1+24j),175(1+24j)}=
\chir^{Vir(3)}_{r,r}.
 \label{lhsexp}
 \een
 This is the first in an infinite series of such identities which
(taking $r=1$, $-2$ and $-5$) rewrite the left hand sides of 
(\ref{id1}).\\

{\bf{Proof of Theorem 5.1} :} Note the congruences $2\ah\equiv 2\mod
8$ and $p\equiv 5\mod 8$ (so also $5p\equiv1\mod 8$). In particular,
$\uh\equiv u\mod p$ and $\uh\equiv1\mod 8$.

\medskip
\noindent
(i) We first rewrite the left hand side of (\ref{dagger}) as a sum of
differences of coset theta functions. 

Now $2m+1=2p(a')^2-1\equiv 2p-1$ modulo $8p$. So by (\ref{VirCons}), 
the $u$th
term in the outer summation on the LHS of $(\ref{dagger})$ is
 $$
\LHS(u)=(\th_{r\uh ,4p}-\th_{r\uh (2p-1),4p})
 (\th_{(2\ah-\e p)r\uh ,4p}-\th_{(2\ah-\e p)r\uh (2p-1),4p}).
 $$
 Applying (\ref{coset}), (with $k=l=h=4p$,
$\lambda=\mu=k_0=\l_0=1$ and $D=-1$)
 \bea
\LHS(u)&=&\th(r\uh\a_1+8p\Okp;16p)-\th(r\uh\a_2+8p\Okp;16p)\nn
 &&\qquad -\th(r\uh\a_3+8p\Okp;16p)+\th(r\uh\a_4+8p\Okp;16p),
\label{star}\eea 
where,
$$ \eqalign{
 \a_1&=1+(2\ah -p\e)i;\cr
 \a_2&=1-(2\ah -p\e)(2p-1)i);\cr
 \a_3&=(2p-1)-(2\ah-p\e)i
    \equiv (2p-1)\a_2\mod 8p;\cr
 \a_4&=(2p-1)(1+(2\ah-p\e)i)
    = (2p-1)\a_1.  \cr} $$
(Note that $(2p-1)^2\equiv1\mod 8p$.)  Now
$\uh $ is $u$ mod $p$ and 1 mod 8. So 
 $$\eqalign{
\uh (2p-1)&\equiv -u \equiv p-u\equiv\widehat{p-u}\ {\rm  mod}\ p\cr
\noalign{\leftline{and}}
\uh (2p-1)&\equiv 1(10-1) \equiv 1\equiv\widehat{p-u}\ {\rm  mod}\ 8\cr
\noalign{\leftline{and therefore}}
\hbox{\hglue1in}\uh (2p-1)&\equiv \widehat{p-u}\ {\rm  mod}\ 8p.\cr}$$
 Thus
$$ 
 r\uh\a_4 \equiv r(\widehat{p-u})\a_1\quad{\rm and} \quad
 r\uh\a_3 \equiv r(\widehat{p-u})\a_2. $$
 Hence the LHS of (\ref{dagger}) may be rearranged as
  \ben
\sum_{u=1}^{\frac{p-1}{2}}{\rm LHS}(u)= \sum_{u=1}^{p-1}
   \left(\th(r\uh\a_1+8p\Okp;16p)-\th(r\uh\a_2+8p\Okp;16p)\right),
 \label{sumstretch}
 \een
 since doubling the range of the sum exactly compensates for the
 elimination
of the last two terms of (\ref{star}). We now prepare to express these 
coset
theta functions as ray class theta functions using (\ref{bridge}) and
to reduce the natural conductors using Theorem 4.4.

 We note first that $p\Okp= P\overline P$, where 
 $P=(1-2\ah i)_\Okp$. So (cf. \ref{primedecomp}), $P$ and
$\overline{P}$ are distinct prime ideals of norm p and by
(\ref{modpiso}),
 \ben
 (\Okp/P)^\times = \{u+P\mid u\in\Zop,\  1\le u\le p-1\}.
\label{unitsmodPd}
 \een

We shall apply Theorem 4.4 with $\f$, $P$, $F$, $B$ and $T$ of that
theorem being respectively $\fp$, $P$, $F'=(P'_2)^{6-\e}$,
$A'_{F'}\,(=\{[1]_{F'}\}$ by \ref{Ad8}) and
$S'_{F'}\,(=\{[\delta]_{F'}\}$ by \ref{skew1}).  The choice of $P$ and
$T$ are consistent since $1-2\ah i\equiv 1-2i\mod 4$ and so, from
(\ref{char2}), $\psi([P])=-1 $.

We express the elements of
${\cal{C}}_{PF'}(\fp)$ in the form $[\a,\b]_{PF'}$ as explained
after (\ref{chinrem}), following the decomposition
 $$ (\Okp/PF')^\times
 \cong(\Okp/P)^\times\times (\Okp/F')^\times,
 $$
 Then, by (\ref{liftofsubgroup}) and (\ref{unitsmodPd}),
 $\tilde B$ of Theorem 4.4 is
  $$ \tilde B\=\left({\rm red}_{F'}^{PF'}\right)^{-1}(B)
 =\{[u,1]_{PF'}\mid 1\le u\le p-1\}
 =\{[\uh ]_{PF'}\mid 1\le u\le p-1\}
 $$
  and $\tilde T=\tilde B[\delta']_{PF'}$, provided
$[\delta']_{F'}=[\delta]_{F'}$.\\

\medskip \noindent (ii) We examine first the case when $\epsilon =0$.
 Then $\a_1\Okp=\overline{P}$ and so, for $\b \in \Okp$ prime
to $2P$, the highest common factor $H$ of
 $(\b \a_1)_{\Okp}$ and
 $(8p)_{\Okp}=8P \overline{P}$ is
 $\overline{P}$. Therefore, using the relation (\ref{bridge})
between coset and ray class theta functions for the ray class group
${\cal{C}}_{PF'}(\fp)$, we get,
  \ben
\theta (\b\a_2+8p\Okp;16p)=\theta([\b]_{PF'};16).
\label{genred1}
\een
 Now $\tilde T=\tilde B[\a_2/\a_1]_{PF'}$,
since $\a_2/\a_1=1-4(1-2ai)i$ is prime to $P$ and congruent to
$\delta$ modulo 8. Thus, applying (\ref{genred1}) to (\ref{sumstretch})
(with $\b=r\uh$ or $r\uh\a_2/\a_1$) we find that the LHS of 
(\ref{dagger}) is
  \bea
 \sum_{u=1}^{p-1}
    \left( \th([r\uh]_{PF'};16)-\th([r\uh\a_2/\a_1]_{PF'};16)\right)
 &=& \th(\tilde B[r]_{PF'},16) - \th(\tilde T[r]_{PF'},16)\cr
 &=& \th([r]_F-[r\delta]_F;16),
 \eea
by Theorem 4.4, as required.

\medskip
\noindent
(iii) Now consider the case when   $\epsilon=1$. 

{} For $\b \in \Okp$ prime
to $2P$, the highest common factor, $H$, of
 $(\b (1+2ai)(1+i))_{\Okp}=\b\overline P P_2$ and
 $(8p)_{\Okp}={P'_2}^6P \overline{P}$ is
 $\overline{P}P'_2$. Therefore, using (\ref{bridge}) again,
we get
  \ben
\theta (\b(1+2ai)(1+i)+8p\Okp;16p)=\theta([\b]_{PF'};8),
\label{genred2}
\een
Put $\b_j= {\a_j}/((1+2\ah i)(1+i))$, for $j=1,\,2$. Then,
since $2a\equiv2\mod8$,
   $$
(1+i)\b_1 = {1+2\ah i-pi\over(1+2\ah i)}
={1-(1-2\ah i)i}\equiv -1-i\mod8.
  $$
 So $\b_1\equiv-1$ modulo $(8/(1+i))_\Okp=F'$ and so $[\b_1]_{F'}
=[1]_{F'}$ and (from the third expression) $\b_1$ is prime to $P$.
 Thus
 $\tilde B[\b_1]_{PF'}=\tilde B$.

Again
   $$
(1+i)\b_2 = {1+2\ah i-ip(1-2p+4a)\over(1+2\ah i)}
={1-i(1-2\ah i)(1-2p+4a)}.
  $$ 
 So $\b_2$ is prime to $P$ and 
$(1+i)\b_2 \equiv 1-(i+2)(1-2+4)\equiv-2-3(1+i)$ modulo 8.
So  
 $$
 \b_2 \equiv -(1-i)-3=-(4+i)\equiv -i\delta\mod 4P'_2. 
 $$
 Hence $\tilde B[\b_2]_{PF'}=\tilde T$.
 Thus, applying (\ref{genred2}) to (\ref{sumstretch})
(with $\b=r\uh\b_1$ or $r\uh\b_2$) we find that the LHS of 
(\ref{dagger}) is
  $$
 \sum_{u=1}^{p-1}
   \left(  \th([r\uh\b_1]_{PF'};8)-\th([r\uh\b_2]_{PF'};8)\right)
 = \th(\tilde B[r]_{PF'},8) - \th(\tilde T[r]_{PF'},8)
 = \th([r]_F-[r\delta]_F;8),
  $$
by Theorem 4.4, as required.

\subsection{The second set of identities}

The proof of the second set of identities (\ref{id2}) relies on
Theorem 4.3, as did the proof of the identities (\ref{id1}) in
subsection 5.1. However now, the relevant quadratic fields are
$\f=\Qop[\rp]$ and $\fp=\Qop[\rpd]$.  By (\ref{algint}), their rings
of algebraic integers are $\Ok = \Zop[\sqrt{-30}]$ and $\Okp =
\Zop[\sqrt{-10}]$.  By (\ref{cond1}, \ref{cond2}), the conductors of
the ray class characters $\psi$ and $\psi'$ introduced in subsection
4.1 are given by,
 \bea
 \eqalign{
F_\psi&=(2^a)_{\Ok}=(2)_{\Ok}=P_2^2\quad\hbox{and }\cr
 F_{\psi'}&=(3\by 2^a)_{\Okp}=(6)_{\Okp}=3{P_2'}^2,\cr}
 \eea
since
$D''=3$ and so $\tilde D''=12$, and $a=1$.

 The calculation of $\psi$ and $\psi'$ is complicated by the fact 
that ideals in $\Ok$ and $\Okp$ need not be principal ideals:\ \ 
${\cal C}(\f)$ is of order 4 and ${\cal C}(\fp)$ is of order 2.  In 
fact
 $[P_{11}]_{\Ok}$ and $[P_{13}]_{\Ok}$ are generators of ${\cal
C}(\f)$ so that, by (\ref{fullreduction}), $[P_{11}]_{(2)}$ and
$[P_{13}]_{(2)}$ together with ${\cal CP}_{(2)}(\f)$ generate ${\cal
C}_{(2)}(\f)$. We find that $\psi=1$ on $[P_{11}]_{(2)}$ and
$[P_{13}]_{(2)}$ and on $[\g]_{(2)}$ if $\g\equiv 1\mod P_2$ and these
values determine $\psi$.  

Similarly, taking $[P'_{13}]_{\Okp}$ as generator
of ${\cal C}(\fp)$,
 we eventually conclude that $\psi'=1$ on $[P'_{13}]_{(6)}$ and on
$[\a]_{(6)}$, for $\a\in \Ok\setminus(P_2\cup3\Okp)$ if either both or
neither of $\a\equiv 1\mod 2$ and $\a\equiv\pm1$ or $\pm\rpd\mod3$ are
satisfied (this is the only way to ensure that the conductor is no
larger than $6\Ok$ --- note that $\Okp/(3)^\by$ is cyclic of order 8
generated by the coset of $\mu' =1+2\rpd$.).  Again these values
determine $\psi'$.

they

We use the admissible pair $(F,F')$ of conductors where $F=P_5P_34P_2$ 
and 
$F'=P_5'(3)4P_2'$. Thus $\CPF$ is the image of 
 \ben
 (\Ok/F)^\by\cong (\Ok/P_5)^\by\by  (\Ok/P_3)^\by\by (\Ok/4P_2)^\by,
 \een
 (using (\ref{chinrem}) twice) and we denote its elements
$[n,m,\a]_F$, accordingly, slightly generalising the notation
introduced before (\ref{unitrel}).  We find,
 $$
  \ker\psi_F=\langle[P_{11}]_F,[P_{13}]_F,[n,m,1+2\a]_F\mid
   n\in\Zop-5\Zop,m\in\Zop-3\Zop\rangle_{\rm gp}
 $$
 (recalling (\ref{modpiso}) we see that we can take $n$ and $m$ to be
integers). The classes \newline
$[n,m,1+2\a]_F^{1-\g}$ turn out to be trivial,
and for $\mu =1+2\rp$, one has, $$
 [P_{11}]^{1-\g}_F=[\mu/11]_F=[11\mu]_F=[1,-1,3\mu]_F,
$$
since $P_{11}^2=\mu\Ok$  and $11^2\equiv1\mod F$. Similarly, 
$$
 [P_{13}]^{1-\g}_F=[(7+2\sqrt{-30})/13]_F=[-1,1,3\mu]_F.
$$
 So, in the
notation of Theorem 4.3,
 \bea\eqalign{
  A_F &=(\ker\psi_F)^{1-\g}
   =\langle[1,-1,3\mu)]_F,[-1,1,3\mu]_F\rangle_{\rm gp}\cr
  &=\{[1]_F,[1,1,-1]_F,[-1,1,3\mu]_F,[-1,1,-3\mu]_F\}\cr}
 \eea
 Also, $[1,1,1+\r]_F^{1-\g}=[1,1,(1+\r)^2(31)^{-1}]_F=[1,1,-3\mu]_F$, 
and 
 \ben
 S_F=A_F[1,1,-3\mu]_F=A_F[-1,1,1]_F.
 \een

 Again $ \ker\psi'_{F'}=B\cup B[1,\mu',1+\rpd]_{F'}$, where
 $$
 B=\langle[P'_{13}]_{F'},[n,\b,1+2\a]_{F'}\mid
   n\in\Zop-5\Zop,\b\equiv\pm1\ {\rm or}\ \pm\rpd\mod3\rangle_{\rm gp}.
 $$
 We find  $[P'_{13}]_{F'}^{1-\g}=[1,\rpd,-1]_{F'}$,
 $[n,\b,1+2\a]_{F'}^{1-\g}=[1,\pm1,1]_{F'}$, and \newline
 $ [1,\mu',1+\rpd]_{F'}^{1-\g}=[1,-\rpd,-3\mu']_{F'}$.
  So
 \bea\eqalign{
  A'_{F'}&=\langle[1,\rpd,-1]_{F'},
              [1,\rpd,-3\mu']_{F'},[1,-1,1]_{F'}\rangle_{\rm gp}\cr
  &=\{[1,\pm1,1]_{F'},[1,\pm\rpd,-1]_{F'},
              [1,\pm1,3\mu']_{F'},[1,\pm\rpd,-3\mu']_{F'}\}.\cr}
 \eea
 Also, $[1,1,1+\rpd]_{F'}^{1-\g}=[1,1,-3\mu']_{F'}$, and
 \ben
 S'_{F'}=A'_{F'}[1,1,-3\mu']_{F'}=A'_{F'}[1,\pm1,-1]_{F'}.
 \een

\bigskip
 Now (because there are ideals of norm 13 in both $\Ok$ and
$\Okp$) there is an ideal $\tilde P$ in $\Okkp$ such that
 $$
 {\cal N}_{\f\fp/\f}(\tilde P)=P_{13}\quad{\rm and}\quad
 {\cal N}_{\f\fp/\fp}(\tilde P)=P'_{13}.
 $$
 Again, let $p$ be a prime such that $p\equiv1\mod12$. Then, by
quadratic reciprocity,
 $$\left(3\over p\right)=\left(p\over 3\right)=1.$$
  So $\Okpp$ has an ideal $P''_p$ of norm $p$. It follows (from the
identities of section 4.2) that
 $$
 {\cal N}_{\f\fp/\f}(P''_p\Okkp)=p\Ok\quad{\rm and}\quad
 {\cal N}_{\f\fp/\fp}(P''_p\Okkp)=p\Okp.
 $$
 Now, if $n$ is prime to 5 and congruent to 1 mod 12 we may choose (by
Dirichlet's theorem) $p\equiv n\mod 120$. Then, taking $I$ of Theorem
4.3 to be $\tilde P P''_p$, we find that
 $$
 [{\cal N}_{\f\fp/\f}(I)]_F=[pP_{13}]_F=[nP_{13}]_F\quad{\rm and}\quad
 [{\cal N}_{\f\fp/\fp}(I)]_{F'}=[pP'_{13}]_{F'}=[nP'_{13}]_{F'}.
 $$
 So that, by Theorem 4.3,
 \ben
 \th([nP_{13}]_F(A_F-S_F);d)=\th([nP'_{13}]_{F'}(A_{F'}-S_{F'});d)
 \een
 (Here sets $A_F$, $S_F$ etc. stand for the sums of their elements.) 

 In particular, let $r\equiv\zeta\mod 4$, where $\zeta=\pm1$; let
$2-s\equiv\zeta r\mod 8$; and let $t\equiv \zeta s\not\equiv0\mod 5$.
Then, choosing $n$ congruent to $s$ (and $\zeta t$) mod 5, to 1 mod 3
and to $\zeta r$ mod 8, we have
 \ben
 \th([P_{13}]_F[s,1,2-s]_F(A_F-S_F);d)
 =\th([P'_{13}]_{F'}[t,1,r]_{F'}(A_{F'}-S_{F'});d),
 \label{rayclassid3010}
 \een
 where we have used the fact that
$[\zeta,1,\zeta]_{F'}=[1,\zeta,1]_{F'}$ lies in $A'_{F'}$.
 Note that we can take
 \ben
 (s,r,t)=\left\{\ 
 \vcenter{
 \halign{(\ \hfill$#$,\ &\hfill$#$,\ &\hfill$#$\ )&\ \ $\ldots$\ \ 
\hfill#\cr
 1&1&1&(i)\cr
 -11&-5&1&(ii)\cr
 -3&-5&13&(iii)\cr
 -7&1&13&(iv)\cr}}
 \right.
 \label{srtvals}
 \een
 \bigskip

We now set about reducing the identities (\ref{id2}) between Virasoro 
characters to identities like
(\ref{rayclassid3010}).
As a first step, we rewrite the former using the V functions defined in
\ref{notation}. This gives,
\bea
\eta [V(1,4) \pm V(11,4)] &=& [V(1,3) \pm V(5,3)][V(2,5)\mp V(8,5)]\nn
\eta [V(-3,4) \pm V(7,4)] &=& [V(1,3) \pm V(5,3)][V(-4,5) \mp V(14,5)]
\label{id24}
\eea
and
\bea
\eta V(2,4) &=& V(2,3)[V(1,5)-V(19,5)]\nn
\eta V(6,4) &=& V(2,3)[V(7,5)-V(13,5)].
\label{id22}\eea
We concentrate on the four identities (\ref{id24}) here, since the last 
two
(\ref{id22}) can be treated in a very similar way. Using the properties 
of 
V functions described in (\ref{properties}, \ref{VVir}), the relations (\ref{id24})
can be expressed in the following compact form,
\ben
 V(1,2)V(s,4)=VV(r,t),
\een
 where
$$VV(r,t)\=V(r,3)V(2t,5)+V(-5r,3)V(32t,5),$$
with $(s,r,t)$ taking the values (\ref{srtvals}). These equations and 
hence the
first four identities in (\ref{id2}) will follow from 
(\ref{rayclassid3010})
when we show that, for the $(s,r,t)$ of (\ref{srtvals}) and with 
$d=240$,
 \ben
 2V(1,2)V(s,4)= \hbox{LHS of (\ref{rayclassid3010})}
 \label{LHSredn}
 \een
 and
\ben
 2VV(r,t) = \hbox{RHS of (\ref{rayclassid3010})}.
 \label{RHSredn}
 \een

\bigskip
We tackle (\ref{LHSredn}) first.
Applying (\ref{coset}) (both signs), we find
 $$
 2\th_{r,6}\th_{s,20}=T(\a)\=\th(\a+J;d)+\th(\bar\a+J;d),
 $$
 where $\a=10r+s\r$, $J=40P_3$ and $d=2400$. Assume $r$ prime to 6 and
$s$ to 10. We find that the h.c.f of $\a\Ok$ and $J$ is $H=P_2P_5$
and, since $F=JH^{-1}$,
 $$
 T(\a)=\th([\a H^{-1}]_F+[\bar\a H^{-1}]_F;240)
      =\th([\a H^{-1}]_F([1]_F+[\bar\a/\a]_F);240).
 $$
 Now put $\b=10+\r$ and, choosing $b$ prime to 10, put
$\hat\a=10r+bs\r$. Then
 $$
 \a\hat\b-\hat\a\b= 10(b-1)(r-s)\r\in  10F.
 $$
 Since $H^2=10\Ok$, it follows from Lemma 3.3(ii) that
 $$
  [\hat\a/\a]_F=[\hat\b/\b]_F=\cdots=[b,1,2-b+(b-1)\r]_F=\e(b),\ 
{\rm say},
 $$
 using the 3-component notation developed above. Also,
$N(\b H^{-1})=13$. So we may take $P_{13}=\b H^{-1}$ and then
$[\a H^{-1}]_F=[P_{13}]_F\e(s)$. We have now
 $$
 T(\a)=\th([P_{13}]_F\e(s)([1]_F+\e(-1));240).
 $$
 Now, since $-31\equiv 2m+1\mod 2m(m+1)$ for $m=2$ and $m=4$,
 (\ref{notation}) can be rewritten for these $m$ values as,
 $$
  V(r,m)=\theta_{r,m(m+1)}-\theta_{31r,m(m+1)}.
 $$
So, taking $b=31$ and $r=1$ in $\a$ and $\hat\a$,
 $$
 2V(1,2)V(s,4)=T(\a)+T(31\a) - T(\hat\a)-T(31\hat\a)=\th(X(s);240)
 $$
 where
 $$
 X(s)=[P_{13}]_F\e(s)([1]_F+[31]_F)([1]_F+\e(-1))([1]_F-\e(31)).
 $$
Now, $[31]_F=[1,1,-1]_F$, $\e(-1)=[-1,1,3\mu]_F$
 and $\e(31)=[1,1,3\mu]\in S_F$. So
 $$
 X(s)=[P_{13}]_F\e(s)(A_F-S_F)=[P_{13}]_F[s,1,2-s]_F(A_F-S_F),
 $$
 if $s\equiv1\mod4$. Thus we have proved (\ref{LHSredn}). 
\bigskip

Applying (\ref{coset}) again, we find
 $$
 2\th_{r,12}\th_{2t,30}=T(\a)\=\th(\a+J;d)+\th(\bar\a+J;d)
 $$
 where $\a=5r+2t\rpd$, $J=60P'_2$ and $d=1200$. 

We assume
 \ben
 \hbox{$r\equiv1\mod 6$ and $t$ prime to 5 and
$t\equiv1\mod3$.}
\label{assumptions}
\een
 We find that the h.c.f of $\a\Ok$ and $J$ is
$H=P'_5$ and, since $F'=JH^{-1}$,
 \ben
 T(\a)=\th([\a H^{-1}]_{F'}+[\bar\a H^{-1}]_{F'};240)
      =\th([\a H^{-1}]_{F'}([1]_{F'}+[\bar\a/\a]_{F'});240)
 \label{Talpha1}
 \een
 Now put $\b=5+2\rpd$ and, choosing $a\equiv1\mod6$ and $b$ prime to 15, 
put
$\hat\a=5ar+2bt\rpd$. Then
 $$
 \a\hat\b-\hat\a\b= 10(b-a)(r-t)\rpd\in 5F',
 $$
since $r\equiv t\mod 3$.
 Since $H^2=5\Ok$, it follows  from Lemma 3.3(ii) that
 \ben
  [\hat\a/\a]_{F'}=[\hat\b/\b]_{F'}=\cdots=[1,1,a]_{F'}\e'(b),
 \label{modquot}
 \een
 where
 $$\eqalign{
 \e'(b)&=[b,-(1+b)-(b-1)\rpd,1+2(b-1)\rpd]_{F'}\cr
       &=[b,1,1]_{F'},\quad \hbox{if $b\equiv1\mod6$.}\cr
 }$$
 In particular, from (\ref{Talpha1}),
 \ben
 T(\a) =\th([\a H^{-1}]_{F'}([1]_{F'}+\e'(-1));240).
 \label{Talpha2}
 \een
 Now, from  (\ref{notation}),
 \ben
  V(r,3)=\th_{r,12}-\th_{7r,12}\quad{\rm and}\quad
  V(2t,5)=\th_{2t,30}-\th_{2(-11t),30},
 \label{notations2}
 \een
 where we took the minus sign so that $7\equiv-11\equiv1\mod 3$. This
ensures that if $r=r_0$ and $t=t_0$ satisfy (\ref{assumptions}) then
so do all the pairs $(r,t)$ of the products
$T(\a)=2\th_{r,12}\th_{2t,30}$ in the expansion of $VV(r_0,t_0)$ using
(\ref{notations2}). Expressing each such product as in (\ref{Talpha2})
and using (\ref{modquot}) several times with different choices of $a$
and $b$, we find, writing $\a_0=5r_0+2t_0\rpd$, that
 \ben
 VV(r_0,t_0)=\th(X;240),
 \label{RHSVV}
 \een
 where
 $$
\eqalign{X=[\a_0 H^{-1}]_{F'}&([1]_{F'}+[1,1,-5]_{F'}\e'(16))\times\cr
 &\times([1]_{F'}-\e'(-11)-[1,1,7]_{F'}+[1,1,7]_{F'}\e'(-11))
 ([1]_{F'}+\e'(-1))\cr
\ldots=[\a_0 H^{-1}]_{F'}&(A_{F'}-S_{F'}).\cr}
 $$
 Now, $N(\b)=25+40=65$. So $N(\b H^{-1})=13$. Hence we may choose
$P'_{13}=\b H^{-1}$ and then
 $$
 [\a_0H^{-1}]_{F'}=[P'_{13}]_{F'}[\a_0/\b]_{F'}
 =[P'_{13}]_{F'}[1,1,r_0]\e(t_0).
 $$
 So, if $t_0\equiv1\mod6$,
 $$
 X=[P'_{13}]_{F'}[t_0,1,r_0]_{F'}(A_{F'}-S_{F'}).
 $$
 Thus we have proved (\ref{RHSredn}) and, as observed there, the first
and second lines of (\ref{id2}) now follow.

\bigskip
\section{Conclusions}

Over the years, two-dimensional conformal field theory has proven to
be a true goldmine for those studying string theory as well as 
statistical mechanics.
Its underlying algebraic structure is the infinite dimensional 
Virasoro algebra. Although its representation theory has been thoroughly analysed, it is remarkable that identities between unitary minimal 
Virasoro characters of low
level, of the kind discussed in this paper, have not been of use in any
`physical' context we are aware of. These identities could therefore 
be regarded as mathematical curiosities, but our aim here has been 
to provide a solid mathematical framework within which they naturally 
appear as the consequence
of relations between two `well chosen' imaginary quadratic extensions over
$\Qop$. The formalism used is borrowed from number theory, and provides, together with a new proof of the identities (\ref{id1},~\ref{id2}), a new
infinite family of identities between Virasoro characters at level 3 and 
level $m=4a^2$, for $a$ odd and $1+4a^2=a'^2p$ where $p$ is prime.

{}From the number theory point of view, the interesting result is 
Theorem 4.3,
which describes relations
between ray class theta functions of two different
imaginary quadratic fields $\f$ and $\fp$, under a certain number 
of constraints presented in Section 4.
That these relations, when the pair ($\f$,$\fp$) is different from ($\Qop[\sqrt{-2}],
\Qop [\sqrt{-1}]$) and ($\Qop[\sqrt{-30}],\Qop [\sqrt{-10}]$) but still 
obeys the constraints of Section 4, lead to other
identities between unitary minimal Virasoro characters, is neither proven
nor disproven at this stage.

\section*{Acknowledgments}
One of us (A.T.) acknowledges the U.K. Engineering and Physical Sciences
Research Council for the award of an Advanced Fellowship.

\end{document}